\documentclass[12pt,titlepage]{utarticle}
\input epsf

\newcount\figno
\def\fig#1#2#3{
\par\begingroup\parindent=0pt\leftskip=1cm\rightskip=1cm\parindent=0pt
\baselineskip=11pt
\global\advance\figno by 1
\epsfxsize=#3
\centerline{\epsfbox{#2}}
\vskip 12pt
{\bf Figure \the\figno:} #1\par
\endgroup\par
}
\def\figlabel#1{\xdef#1{\the\figno}}
\def\encadremath#1{\vbox{\hrule\hbox{\vrule\kern8pt\vbox{\kern8pt
\hbox{$\displaystyle #1$}\kern8pt}
\kern8pt\vrule}\hrule}}

\newcommand\semidirect{\mathbin{\hbox{\hskip2pt\vrule height 5.0pt depth -.3pt
width .25pt \hskip-2pt$\times$}}}
\newcommand{\bi}{\bibitem}
\newdimen\tableauside\tableauside=1.0ex
\newdimen\tableaurule\tableaurule=0.4pt
\newdimen\tableaustep
\def\phantomhrule#1{\hbox{\vbox to0pt{\hrule height\tableaurule width#1\vs
s}}}
\def\phantomvrule#1{\vbox{\hbox to0pt{\vrule width\tableaurule height#1\hs
s}}}
\def\sqr{\vbox{%
  \phantomhrule\tableaustep
  \hbox{\phantomvrule\tableaustep\kern\tableaustep\phantomvrule\tableauste
p}%
  \hbox{\vbox{\phantomhrule\tableauside}\kern-\tableaurule}}}
\def\squares#1{\hbox{\count0=#1\noindent\loop\sqr
  \advance\count0 by-1 \ifnum\count0>0\repeat}}
\def\tableau#1{\vcenter{\offinterlineskip
  \tableaustep=\tableauside\advance\tableaustep by-\tableaurule
  \kern\normallineskip\hbox
    {\kern\normallineskip\vbox
      {\gettableau#1 0 }%
     \kern\normallineskip\kern\tableaurule}%
  \kern\normallineskip\kern\tableaurule}}
\def\gettableau#1 {\ifnum#1=0\let\next=\null\else
  \squares{#1}\let\next=\gettableau\fi\next}

\newcommand{\fQn}[1]{Q^{#1}}
\newcommand{\aQn}[1]{\ot{Q}_{#1}}
\newcommand{\ot}[1]{\bar{#1}}
\newcommand{\bea}{\begin{eqnarray}}
\newcommand{\eea}{\end{eqnarray}}
\newcommand{\be}{\begin{equation}}
\newcommand{\ee}{\end{equation}}
\newcommand{\nn}{\nonumber}

\newcommand{\complex}{{{\rm I} \kern -.59em {\rm C}}}
\def\slashD{D\hskip-0.75 em / \hskip+0.30 em}
\def\Nf{{N_f}}
\def\qp{{q'}}
\def\adot{{\dot{\alpha}}}
\def\bdot{{\dot{\beta}}}

\def\I{{\scriptscriptstyle I}}
\def\a{\alpha}
\def\qbar{{\bar{q}}}
\def\fund{ \> {\vcenter  {\vbox
	    {\hrule height.6pt
	 \hbox {\vrule width.6pt  height5pt
	  \kern5pt
		 \vrule width.6pt  height5pt }
	\hrule height.6pt}}} \>\> }
\def\xrm{{\rm \X}}
\def\fund{  \> {\vcenter  {\vbox
              {\hrule height.6pt
               \hbox {\vrule width.6pt  height5pt
                      \kern5pt
                      \vrule width.6pt  height5pt }
               \hrule height.6pt}
                         }
                   }
           \>\> }
\def\antifund{  \> \overline{ {\vcenter  {\vbox
              {\hrule height.6pt
               \hbox {\vrule width.6pt  height5pt
                      \kern5pt
                      \vrule width.6pt  height5pt }
               \hrule height.6pt}
                         }
                   } }
           \>\> }

\def\sym{  \> {\vcenter  {\vbox
              {\hrule height.6pt
               \hbox {\vrule width.6pt  height5pt
                      \kern5pt
                      \vrule width.6pt  height5pt
                      \kern5pt
                      \vrule width.6pt height5pt}
               \hrule height.6pt}
                         }
              }
           \>\> }

\def\symbar{  \> \overline{ {\vcenter  {\vbox
              {\hrule height.6pt
               \hbox {\vrule width.6pt  height5pt
                      \kern5pt
                      \vrule width.6pt  height5pt
                      \kern5pt
                      \vrule width.6pt height5pt}
               \hrule height.6pt}
                         }
              }
           } \>\> }
\def\anti{ \> {\vcenter  {\vbox
              {\hrule height.6pt
               \hbox {\vrule width.6pt  height5pt
                      \kern5pt
                      \vrule width.6pt  height5pt }
               \hrule height.6pt
               \hbox {\vrule width.6pt  height5pt
                      \kern5pt
                      \vrule width.6pt  height5pt }
               \hrule height.6pt}
                         }
              }
           \>\> }

\def\antithree{ \>
              {\vcenter  {\vbox
              {\hrule height.6pt
               \hbox {\vrule width.6pt  height5pt
                      \kern5pt
                      \vrule width.6pt  height5pt }
               \hrule height.6pt
               \hbox {\vrule width.6pt  height5pt
                      \kern5pt
                      \vrule width.6pt  height5pt }
               \hrule height.6pt
               \hbox {\vrule width.6pt  height5pt
                      \kern5pt
                      \vrule width.6pt  height5pt }
               \hrule height.6pt}
                         }
              }
           \>\> }\def\antifour{ \>
              {\vcenter  {\vbox
              {\hrule height.6pt
               \hbox {\vrule width.6pt  height5pt
                      \kern5pt
                      \vrule width.6pt  height5pt }
               \hrule height.6pt
               \hbox {\vrule width.6pt  height5pt
                      \kern5pt
                      \vrule width.6pt  height5pt }
               \hrule height.6pt
               \hbox {\vrule width.6pt  height5pt
                      \kern5pt
                      \vrule width.6pt  height5pt }
               \hrule height.6pt
               \hbox {\vrule width.6pt  height5pt
                      \kern5pt
                      \vrule width.6pt  height5pt }
               \hrule height.6pt}
                         }
              }
           \>\> }
\def\antifive{ \>
              {\vcenter  {\vbox
              {\hrule height.6pt
               \hbox {\vrule width.6pt  height5pt
                      \kern5pt
                      \vrule width.6pt  height5pt }
               \hrule height.6pt
               \hbox {\vrule width.6pt  height5pt
                      \kern5pt
                      \vrule width.6pt  height5pt }
               \hrule height.6pt
               \hbox {\vrule width.6pt  height5pt
                      \kern5pt
                      \vrule width.6pt  height5pt }
               \hrule height.6pt
               \hbox {\vrule width.6pt  height5pt
                      \kern5pt
                      \vrule width.6pt  height5pt }
               \hrule height.6pt
               \hbox {\vrule width.6pt  height5pt
                      \kern5pt
                      \vrule width.6pt  height5pt }
               \hrule height.6pt}
                         }
              }
           \>\> }

\def\twotwo{
              {\vcenter  {\vbox
              {\hrule height.5pt
               \hbox {\vrule width.5pt  height4pt
                      \kern4pt
                      \vrule width.5pt  height4pt
                      \kern4pt
                      \vrule width.5pt height4pt}
               \hrule height.5pt
               \hbox {\vrule width.5pt  height4pt
                      \kern4pt
                      \vrule width.5pt  height4pt
                      \kern4pt
                      \vrule width.5pt height4pt}
               \hrule height.5pt}
                         }
              }
           \>\> }

\def\four{  {\vcenter  {\vbox
              {\hrule height.5pt
               \hbox {\vrule width.5pt  height4pt
                      \kern4pt
                      \vrule width.5pt  height4pt
                      \kern4pt
                      \vrule width.5pt  height4pt
                      \kern4pt
                      \vrule width.5pt  height4pt
                      \kern4pt
                      \vrule width.5pt height4pt}
               \hrule height.5pt}
                         }
              }
           }
\def\Vslash{V\hskip-0.75 em / \hskip+0.30 em}
\def\Rtwo{ {\Nf-4 \over \Nf+4} }
\def\A{{\scriptscriptstyle A}}

\def\I{{\scriptscriptstyle I}}
\def\J{{\scriptscriptstyle J}}
\def\K{{\scriptscriptstyle K}}
\def\L{{\scriptscriptstyle L}}

\def\S{{\scriptscriptstyle S}}
\def\T{{\scriptscriptstyle T}}

\def\X{{\scriptscriptstyle X}}

\def\a{\alpha}
\def\b{\beta}

\def\e{\epsilon}

\def\s{\sigma}

\begin{document}

\preprint{
 HU-EP-97/86\\
 {\tt hep-th/9711157}\\
}

\title{Dualities in all order finite $N=1$ gauge theories}

\author{Andreas Karch
 \thanks{Work supported by the DFG} $\,$, 
 Dieter L\"ust
and George Zoupanos
 \thanks{On leave from the
  Physics Department,
 National Technical University,
 Zografou Campus,
 GR-15780 Athens, Greece.
 Work partially supported by the E.C. projects
FMBI-CT96-1212; ERBFMRXCT960090 and the Greek projects,
PENED95/1170;1981.}
 \oneaddress{
  Humboldt-Universit\"at zu Berlin\\
  Institut f\"ur Physik\\
  Invalidenstra{\ss}e 110\\
  D-10115 Berlin, Germany\\
  {~}\\
  \email{karch@qft1.physik.hu-berlin.de\\
         luest@qft1.physik.hu-berlin.de\\
          george.zoupanos@cern.ch}\\
 }
}

\date{November 20, 1997}

\Abstract{
We search for dual gauge theories of all-loop finite, $N = 1$ supersymmetric
gauge theories. It is shown how to find explicitly the dual gauge theories
of almost all chiral, $N = 1$, all-loop finite gauge theories, while several
models have been discussed in detail, including a realistic finite
$SU(5)$ unified theory. Out of our search only one all-loop, $N = 1$ finite
$SO(10)$ theory emerges, so far, as a candidate for exhibiting also
S-duality.
}

\maketitle

\renewcommand{\baselinestretch}{1.5}
\small\normalsize

\section{Introduction}

 Finiteness is an essential conceptual ingredient in the various
theoretical frameworks that hopefully will lead to a deeper understanding
of nature. There exist many arguments to
believe that the divergencies of ordinary field theory are
not of fundamental nature but rather they are the manifestation of the
existence of new physics at higher scales. Therefore when unification
of all interactions has been achieved the theory might be completely
finite. This is one of the main motivations and aims of
string theory, of non-commutative geometry 
and of other approaches, which try
to include also gravity in the unification of the interactions. 

It is well established that there exist finite theories
without the
inclusion of gravity, i.e. just with gauge interactions. 
For example it is well known
that all $N=4$ and some $N=2$ supersymmetric gauge theories are free from
ultraviolet divergencies at all orders in perturbation theory. Of
particular interest is the fact that there exist $N=1$ supersymmetric gauge
theories, which are finite to all orders in perturbation 
theory \cite{west}.
The 1-
and 2-loop finiteness of these theories is guaranteed by, first, choosing
the particle content such that the 1-loop 
gauge $\beta$-function vanishes, and, second, by
adding a superpotential such that all 1-loop matter field 
anomalous dimensions are zero.
Moreover, the all-loop finiteness requires that the relations between
the Yukawa and gauge couplings, obtained by imposing the
vanishing of the  1-loop anomalous dimensions, should be unique
solutions of the reduction equations, i.e. to be possible to be 
uniquely determined to all orders.
It is worth noting that along these lines
there exist a realistic finite $SU(5)$ GUT which successfully has
predicted, among others, the top quark mass \cite{kmz}.

The discussion about finite $N=1$ gauge theories was so far limited to 
perturbative aspects,
but non-perturbative problems like the bound state spectrum of these 
theories were not
investigated. However,
during the recent years a lot of progress in the understanding 
of supersymmetric gauge theories at strong coupling was achieved. The basic
principle that allows to address non-perturbative
problems is electric-magnetic, strong-weak
coupling duality. $N=4$ supersymmetric gauge theories are now believed to
exhibit an exact Montonen-Olive duality \cite{monol}.
The low-energy effective action of $N=2$ supersymmetric theories \cite{SW}
can be even solved using duality and holomorphy. $N=1$ supersymmetric
gauge theories, which have richer dynamics and are much better
candidates to describe the real world, as compared to the others, exhibit
a weaker version of the electric-magnetic  duality symmetry \cite{seiberg},
 namely gauge theories differing in
the ultraviolet can establish a universality class, since they flow in the
infrared to the same interacting fixed point of the renormalization group.
This means that two different, but dual $N=1$ supersymmetric field theories
with different gauge groups and different numbers of interacting 
particles nevertheless
describe
the same physics at low energies.

The main point in the present paper is to find in a 
systematic way duals of $N=1$ finite
theories in the sense just described.  
Our search covers most of the
chiral $N=1$ finite gauge theories
including the realistic finite $SU(5)$ model.
In the simplest case, for a finite vectorlike $N=1$ $SU(3)$
gauge theory with $N_f=3N_c=9$ fundamental plus antifundamental
quark superfields, the magnetic dual
model is given by a non-finite, asymptotically free $N=1$ model with
gauge group $SU(6)$, and again with $9$ fundamental
matter fields \cite{leigh}. However for more involved models 
with differend kind
of gauge groups and several
types of matter representaion the 
structure of the dual magnetic models is apriori not determined. 
Indeed, we will
find that the dual of a finite $N=1$ theory is almost never finite, but it
will be either asymptotically
free or non-free. However for one particular model, namely 
$SO(10)$ with matter fields
in $N_f=8$ vector and $N_q=8$ spinor representations, the dual theory 
has also vanishing
one-loop $\beta$-function. Adding a superpotential to this model, 
both the electric
as well as the magnetic theories are hoped to be all order finite. 
Therefore this model
may serve as the first
step in searching for
$S$-duality in $N=1$ theories analogous to the one appearing in $N=4$ theories.

The presence of the duality between seemingly unrelated theories most 
probably indicates that
they are parts of a more fundamental underlying theory. Indeed, 
studying non-perturbative duality symmetries in 
string theory \cite{stringdual},
many of the fields theory dualities find a very beautiful, 
geometrical explanation.
In particular, many (non)-perturbative problems in gauge theories can be studied
by considering the world volume dynamics of various intersecting 
p-brane configurations,
in the low-energy limit, where the gravitational 
fluctuations are frozen, and only
the low-dimensional
gauge degrees of freedom on the
world volume are relevant \cite{HW}.
In this context, it has been proposed
that configurations of branes in type II string theory are a natural
framework to study four dimensional N=1 supersymmetric gauge dynamics
\cite{EGK}.
The N=1 duality relates theories which in the brane construction can be
connected by a continuous path of rearranging the
brane configurations
in the moduli space of vacua, along which
the gauge symmetry is completely broken and the infrared dynamics is
weak. We examine several examples of brane constructions in which the
resulting low energy field theory is a finite gauge theory. 

Our paper is organized as follows. In the next section we will review
those aspects
of the $N=1$ duality of \cite{seiberg} which will be relevant 
for our further study. Next,
in section three, we will recall the all order finiteness theorems for $N=1$
supersymmetric gauge theories. In section four we will search for 
duals of finite
$N=1$, 
$SU(N_c)$ gauge theories, namely first for a simple $SU(3)$ gauge 
group with $9$
vectorlike matter fields \cite{leigh}, 
second for vectorlike finite $SU(N_c)$ models
with two additional adjoint matter fields and
third for chiral finite $SU(N_c)$ gauge
theories with one symmetric, one antisymmetric plus one adjoint matter field,
where in addition a special type of superpotential is required 
\cite{Brodie+Strassler,TensorKutasov}.
In section five we investigate
an $SO(10)$ model with eight vector and 
spinor matter fields \cite{so10},
where both the initial as well as the dual magnetic theory are 
supposed to be all
order finite.
In section six we will extend the search for duals of finite 
models by using the
deconfinement method \cite{berkooz}, which allows us to generically include also 
an arbitrary number of matter fields which
transform under tensorial representations of the original theory, 
namely symmetric
and antisymmetric tensors of $SU(N_c)$. 
As a particular example within this class we are
able to construct the dual of a realistic $SU(5)$ finite gauge theory.
Finally, in section seven, we will construct finite $N=1$ 
gauge theories and their
duals from brane configurations.

\section{Dualities in $N=1$ gauge theories}

\subsection{Supersymmetric QCD}

Consider the
dynamics of supersymmetric QCD (SQCD) 
with $N_f$ flavors of quarks in the $(N_c+\bar N_c)$ representation 
of the gauge
group \cite{seirev}. The corresponding chiral superfields will be denoted by
$Q_i$ and $\bar Q_i$, $i=1, \dots , N_f$.\footnote{Antichiral 
superfields will be denoted
by dagger.}
So lets start with the case $N_f < N_c$.
At the classical level the global symmetry group is

\be
G_f = SU(N_f)_L \times SU(N_f)_R \times U(1)_B \times U(1)_A \times U(1)_R,
\ee
where the indices $B,A,R$ in the $U(1)$'s refer to baryon number, axial
flavour and axial-$R$ symmetries. All $U(1)$'s are conserved classically but
the two axial $U(1)$'s are broken quantum mechanically due to
anomalies. However the two anomalous $U(1)$'s can be combined to form an
anomaly free $R$-symmetry, say $R_{AF}$, whose charge is
\be
\label{AF}
      Q_{AF} = Q_R +Q_A (N_f - N_c)/N_f 
\ee
with $Q_A$, $Q_R$ refering to the two anomalous $U(1)$ transformations.

In the absence of mass terms, there is a classical moduli space of
vacua \footnote{Every time there exists a continuous family of
solutions to the conditions for unbroken supersymmetry, there exist a
manifold of degenerate vacuum states with zero energy. This manifold will
be typically be parametrized by the vevs of chiral superfields, i.e. it
will be a complex K\"ahler manifold, which is usually called moduli space.}
which can be described in a gauge invariant way by the expectation values
of meson superfields $M_{ij} = Q_i \bar {Q}_j$. The low energy dynamics of SQCD
can be represented by an effective Lagrangian which is built out of the
gauge invariant chiral superfields and would be a generalization of the
corresponding  effective Lagrangian of ordinary QCD. If this Lagrangian is
built out of gauge invariant combinations of $Q_i$ and $\bar{Q_i}$, it must be
a function of $M_{ij}$. The non-perturbative superpotential of this 
effective Lagrangian can
be constructed \cite{ADS, Yank}
 by searching for a function of M which is invariant under
the non-anomalous global symmetry group (and has charge 2 under $Q_{AF}$).
Factors of $\Lambda$ should then be supplied to provide the effective
superpotential with dimension 3. It is
\be
  W_{\rm eff} = c (\Lambda^{(3N_c-N_f)}/\det M)^{1/(N_c-N_f)}.
\label{supo} 
\ee
This superpotential can be further constrained by considering various
limits. For instance for large $M_{ij}$ it can be shown that the gauge
symmetry is spontaneously broken
\be
  SU(N_c) \longrightarrow SU(N_c-N_f).
\ee
Probably the most important characteristic of the superpotential is that
it leads to a squark potential which tends to zero as $\det M$ tends to
infinity. Therefore, {\it the quantum theory does not have a ground state}.

 The above superpotentials are linked to each other by holomorphic
decoupling. Then, if this superpotential is known explicitly for a
particular $N_f$, one can compute its coefficient for all $N_f$, $N_c$.
In fact in \cite{ADS} it was shown that there is a direct
derivation of the superpotential for the case $N_f = N_c - 1$. In this case
the vevs of $Q_i$ and $\bar{Q_i}$ break the $SU(N_c)$ gauge symmetry
completely.

 The next step is to see what happens for larger values of $N_f$. Let us
start with the case $N_f = N_c$. The naive intuition that it would just be a
smooth extrapolation of the previous cases fails. Clearly the 
superpotential (\ref{supo}) is singular in this case. However a
very interesting new feature now is the fact that is the first
case that it is possible to build gauge invariant chiral fields with the
quantum numbers of baryons. There exist two such terms
\be
B = \epsilon_{j_1\dots j_{N_c}}\epsilon_{\alpha_1 ... \alpha_{N_c}}
Q_{j_1}^{\alpha_1} ... Q_{j_{N_c}}^{\alpha_{N_c}},
\ee
\be
\bar{B} = \epsilon_{j_1\dots j_{N_c}}\epsilon_{\alpha_1...\alpha_{N_c}}
\bar{Q}_{j_1}^{\alpha_1}...\bar{Q}_{j_{N_c}}^{\alpha_{N_c}},
\ee
where the down, up indices denote flavour, colour respectively.
The classical moduli space has again a gauge invariant description in
terms of the vevs of mesons $M_{ij}$ and baryons $B$
and $\bar{B}$, subject to
the classical constraint
\be
  \det M - \bar{B} B = 0,
\label{clcon}
\ee
which follows from Bose statistics of $B$, $\bar{B}$.
One might think that the low enery dynamics of this theory is being
described by the fields $M$, $B$, and $\bar{B}$
 which are fluctuating subject to
the constraint (\ref{clcon}). However it was argued \cite{Se}
 that this manifold of
vacua is distorted by non-perturbative effects. In fact there is no
symmetry which prohibits the modification of the constraint 
(\ref{clcon}) to
\be
\label{qcon}
 \det M - \bar{B} B = \Lambda^{2N_c} ,
\ee
as was argued in \cite{Se}.

 The case with $N_f = N_c$ provides a first example of a theory with a
moduli space (manifold of vacuum states). Note that the origin $M = B
= \bar{B} = 0$ is not on the moduli space due to (\ref{qcon}).
This in turn implies
that the quantum dynamics necessarily break the anomaly free chiral
symmetry. In fact different points on the quantum moduli space exhibit
different patterns of chiral symmetry breaking. Moreover since there are
no singularities on the quantum moduli space, the only massless particles
are the moduli, the fluctuations of $M$, $B$, $\bar{B}$, preserving 
(\ref{qcon}). In the
semiclassical limit it is appropriate to think of the theory as being in
the Higgs phase. On the other hand near the origin, since the theory is
smooth in terms of mesons and baryons, it is appropriate to think of the
theory as being in the confining phase. There is a smooth transition among
the two regions.

Since the spectrum contains massless composite fermions one has to check whether
't Hoofts's anomaly matching conditions \cite{hooft} are satisfied.
 At a generic point in the moduli space, 
the original global symmetry of the model
eq.(1) (note that $U(1)_R$ and $U(1)_{AF}$ coincide in this case)
is broken down to $U(1)_R$ by the vevs of the fields $M$,
$B$
and $\bar{B}$. However there are certain special points of maximal
symmetry, where a large part of $G$ remains unbroken.
At these points, the 't Hooft's conditions are especially strong.
It is very encouraging that the conditions can be satisfied in the cases
that $G$ is broken to
\be
 SU(N_f)_V \times U(1)_B \times U(1)_R
\ee
and to
\be
 SU(N_f)_L \times SU(N_f)_R \times U(1)_R.
\ee
 Therefore the above picture of the behaviour of SQCD for $ N_f = N_c $
passes a highly non trivial consistency check.

 With the above experience one can go on and discuss the case with $N_f =
N_c + 1$. The classical moduli is again described by the gauge invariant
superfields mesons $M$ and baryons
\be
B_i = \epsilon_{ij_1 ... j_{N_c}} \epsilon_{\alpha_1...\alpha_{N_c}}
Q_{j_1}^{\alpha_1}...Q_{j_{N_c}}^{\alpha_{N_c}},
\ee
\be
\bar{B_i} = \epsilon_{ij_1...j_{N_c}} \epsilon_{\alpha_1...\alpha_{N_c}}
 \bar Q_{j_1}^{\alpha_1}...\bar Q_{j_{N_c}}^{\alpha_{N_c}},
\ee
where the $j_i$, $\alpha_i$ are flavour, colour indices
respectively. The fields $B_i$, $\bar{B_i}$ transform according
to $(\bar{N_f},1)$, $(1,N_f)$, respectively, under $SU(N_f) \times SU(N_f)$. For
this case it was proposed \cite{Se} that the system is described by the
superpotential
\be
 W = 1/\Lambda^{2N_c-1}(\det M - B_iM^{ij}\tilde{B_j}),
\ee
which is invariant under the global symmetry of the model and has charge 2
under the anomaly free $R$-symmetry.

 Some of the characteristic features of this case are the following. Unlike
the previous case here the quantum moduli space is the same as the
classical one
\cite{Se}. The spectrum at the origin of field space consists of
massless composite mesons and baryons and the chiral symmetry of the
theory is unbroken there. Therefore there is confinement without chiral
symmetry breaking. Again there exists a smooth transition among the Higgs
and confining phases. Finally, the 't Hooft's conditions are again met.

\subsection{Seiberg's Duality}
 The picture described above has an obvious generalization in for higher
values of $N_f$. The gauge invariant chiral fields of the theory
are the mesons $M_{ij}$ and the
baryons $B_{ij...k}$, $\bar{B}_{ij...k}$.
The $B_{ij...k}$ is built as a product
of $N_c$ quark superfields which contain all flavours except $i,j,...,k$
and $\bar{B}$ is defined similarly. An $SU(N_f) \times SU(N_f)$ invariant
superpotential is given by
\be
 W \sim (\det M - B_{ij...k} M^{i\bar{i}} M^{j\bar{j}}...M^{k\bar{k}}
\bar{B}_{\bar{i}\bar{j}...\bar{k}}).
\label{mightbe}
\ee
However this superpotential does not have $R$ charge 2, and the  multiplet
of fields $M$, $B$, $\bar{B}$ does not satisfy the 't Hooft's anomaly
conditions. In fact the mismatch grows with each successive number of
flavours. 

To solve this puzzle, observe that the baryon superfields in (\ref{mightbe}) 
have
\be
      \tilde{N}_c = N_f - N_c
\ee
uncontracted indices. 
Making this observation, Seiberg's idea \cite{seiberg} 
now is to regard these fields as
bound states of new superfields  $q_i$ and $\bar{q}_i$, 
$i=1,\dots ,N_f$, which 
transform under the dual gauge group $SU(\tilde N_c)$ 
according to the fundamental, antifundamental
representations respectively. Then the baryon superfields would have the
dual description
\be
 B_{ij...k} = \epsilon_{\alpha_1...\alpha_{\tilde{N}_c}}
q_i^{\alpha_1}q_j^{\alpha_2}...q_k^{\alpha_{\tilde{N}_c}}
\ee
and similarly for $\bar{B}$.

 The complete proposal of Seiberg is that SQCD with gauge group
$SU(N_c)$ and with $N_f$ flavours
for $N_f > N_c + 1$, is dual to,  i.e. can be described by,
 a supersymmetric gauge theory
based on the group $SU(\tilde{N}_c)$ coupled to the elementary chiral
superfields $q_i$, $\bar{q_i}$, $i = 1,...N_f$, as well as to an 
elementary additional
superfield $M^{ij}$, which is gauge singlet. The $M$ couples to $q$, 
$\bar{q}$ via
the tree level superpotential
\be
       \tilde{W} = q M \bar{q}.
\ee
Without the superpotential the theory would have an additional $U(1)$
global symmetry acting on $M$. One can explicitly check that the
superpotential preserves the anomaly free $R$-symmetry. Therefore the newly
constructed theory has the same global symmetry as the original
SQCD. Seiberg \cite{seiberg}
refers to the relation among this theory and the original as
non-Abelian electric-magnetic duality.

\subsection{Fixed points and superconformal invariance}

Consider the exact $\beta$-function\footnote{For a more 
detailed discussion on the exact beta-function
see section 3.2.}
of SQCD with gauge group $SU(N_c)$ 
coupled to
$N_f$ fundamental plus antifundamental matter fields:
\be
\beta(g)=-{g^3\over 16\pi^2}{3N_c-N_f+N_f\gamma(g^2)\over 
1-N_c{g^2\over 8\pi^2}},
\ee
\be
\gamma(g^2)=-{g^2\over 8\pi^2}{N_c^2-1\over N_c}+O(g^4).
\ee
There is a non-trivial zero of the $\beta$-function for $N_f=(3-\epsilon)N_c$,
$N_f,N_c>>1$. In this regime the $\beta$-function becomes
\be
\beta(g)=-{g^3\over 16\pi^2}{\epsilon-{g^2\over 8\pi^2}N_cN_f\over 
1-N_c{g^2\over 8\pi^2}}.
\ee
Therefore, at order $\epsilon$ the fixed point $g^2_*$ is given by
\be
N_cg^2_*={8\pi^2\over 3}\epsilon.
\ee
In fact it was argued in \cite{seiberg} that such a fixed point exists
in the range ${3\over 2}N_c\leq N_f\leq 3N_c$.\footnote{Such a behavior was
already conjectured to hold in ordinary QCD by \cite{BaZa}.
For a recent discussion on the status of this
speculation see \cite{KuCaSt}
and for recent findings in lattice studies
see e.g. \cite{Iwas}.}
In SQCD the key observation is that the superconformal theory at the 
fixed point has a dual
(magnetic) description in terms of a different gauge theory, $SU(N_f-N_c)$.

Let us discuss  a little bit more the structure of the superconformal theory
at the fixed point.
Recall that in supersymmetric theories,
there exists a vector supermultilpet of gauge invariant operators, called
supercurrent \cite{fz-npb87} $J_{sup}$, which contains the classically
conserved currents associated with $R$-invariance, supersymmetry and
translation invariance, i.e. the $R$-current $J_R^{\mu}$, the
supersymmetry current $Q_{\alpha}^{\mu}$ and the energy-momentum
tensor $T^{\mu\nu}$, respectively. It is worth noting that the two latter
currents remain conserved to all orders contrary to the first one. Also
the $R$-current $J_R^{\mu}$ must be the current of the anomaly free
$R$-symmetry, while the anomalous is the one which obtains quantum
corrections only at one loop according to the nonrenormalization
theorem \cite{ab-theo},
 which holds also in the supersymmetric case \cite{pisi}.

 There exist a second, chiral supermultiplet, containing,
among others, the anomalies of the $R$-current $J_R^{\mu}$ as well as the
trace anomalies of the supersymmetry current $Q_{\alpha}^{\mu}$
and the energy-momentum
tensor $T^{\mu}_{\nu}$.
This supermultiplet is called supertrace anomaly and can be proven
that is proportional to the full $\beta$-function $\beta_g$. Therefore at the
infrared fixed point of SQCD it holds that
\be
 T_{\mu}^{\mu} = 0 , \; \; \sigma_{\alpha \dot{\beta}}^{\mu} 
\bar{Q}_{\mu}^{\dot{\beta}} = 0, \;
\; \partial_{\mu}J_R^{\mu} = 0.
\ee
The superconformal algebra gives restrictions on the eigenvalues of these
operators. In particular the scaling dimension of a field is bounded by
its $R$ charge
\be
  d \geq 3/2|R| ,
\ee
which holds strictly for gauge invariant operators and the inequality is
saturated for chiral and antichiral superfields
\cite{sohnius}.
This has important consequences.
For instance consider the operator product of two chiral
operators $O_1(x)O_2(0)$. All the the operators in the resulting
expansion have $R = R(O_1) + R(O_2)$ and therefore $d \geq d(O_1) + d(O_2)$.
Thus there is no singularity in the expansion at $x = 0$ and we can
define the product of the two operators by simply taking the limit of $x$
to zero. If this limit does not vanish, it  leads to a new chiral
operator $O_3$ whose dimension is $d(O_3) = d(O_1) + d(O_2)$.
Since the basic
objects of the description so far are chiral superfields, one can work out
their scaling dimensions from their R charge. In particular
\be
 Q \bar{Q} = M \; \; \mbox{  has  } \; \;
  d = {3\over 2}R(Q \bar{Q}) = {3\over 2}(N_f - N_c)/N_f 
\ee
\be
 q \bar{q} = U \; \; \mbox{   has   } \; \;
 d = {3\over 2}R(q\bar{q}) = 3(N_c/N_f) .
\ee
In addition the superpotential $\tilde{W}
 = q M \bar{q}$ has $R = 2$, as it should
in order to preserve $R$-symmetry and $d = 3$, which is the correct value
for a marginal perturbation.
In SQCD  the one loop coefficient of
the $\beta$-function vanishes at $N_f = 3N_c$.
At this point, the bilinear $U$
has $d = 1$ which is the dimension of a free field. Similarly, the dimension
of the bilinear $M$ reaches 1 at $N_f = 3N_c/2$.
This value of $N_f$ coincides
with the value in which the one loop coefficient of the dual
$\beta$-function vanishes and below which the dual theory becomes
asymptotically non-free.

In summary, the behavior of SQCD with gauge group $SU(N_c)$ and
$N_f$ fundamental plus antifundamental matter fields as a function of the 
parameter $N_f$
can be characterized as follows. The electric 
coupling becomes stronger for smaller $N_f$, 
whereas the magnetic coupling becomes stronger for larger $N_f$.
Within the range ${3\over 2}N_c\leq N_f\leq 3N_c$ there is a 
superconformal fixed
point for finite coupling constant. Under 
Seiberg's duality this range is mapped to
itself with self dual point
$N_f=2N_c$;  an 
electric theory with $N_f=3N_c$ and vanishing
one-loop $\beta$-function gets mapped to an asymptotically
free magnetic theory with $\tilde N_f={3\over 2}\tilde N_c$, and 
vice versa. We will see
in the following that for general gauge groups with several 
matter representations
the magnetical (electric) dual of a finite
 electric (magnetic) theory is not
necessarily asymptotically free, but can by 
either asymptotically free or asymptotically non-free or,
in a particular case, again finite, i.e. has vanishing $\beta$-function.

In our search in the present paper we shall examine theories exactly
at the point in which the one loop gauge $\beta$ function vanishes. Moreover
as we shall see the anomalous dimensions of the superfields will also
vanish at one loop
which in turn will permit us to meet the conditions necessary to
have vanishing $\beta$-functions to all loops. 
Since we will consider $SU(N_c)$ gauge theories not only with matter fields
in the fundamental representation, and also other gauge groups than $SU(N_c)$,
let us give the general expression for the one-loop $\beta$-function in terms of
$\mu_{\rm gauge}=C_2(G)$ and $\mu_{\rm matter}=\sum_il(R_i)$, where 
$C_2(G)$ is the quadratic 
Casimir of the adjoint representation of the gauge group $G$ and 
$l(R_i)$ is the Dynkin
index of the matter representation $R_i$ \cite{PW}:
\be
\beta_g^{(1)}=
\frac{d g}{d t} =
-{g^3\over 16\pi^2}(3\mu_{\rm gauge}-\mu_{\rm matter})=
\frac{g^3}{16\pi^2}\,[\,\sum_{i}\,l(R_{i})-3\,C_{2}(G)\,]~.
\label{betag}
\ee
For $\mu_{\rm matter}<\mu_{\rm gauge}$ there is an effective 
superpotential analogous
to eq.(3), and at the point $\mu_{\rm matter}=\mu_{\rm gauge}$ 
there is a quantum smoothed
moduli space of the form eq.(8). For $\mu_{\rm matter}<3\mu_{\rm gauge}$ 
($\mu_{\rm matter}>3\mu_{\rm gauge}$) the theory is
asymptotically (infrared) free, and at 
$\mu_{\rm matter}=\mu_{\rm gauge}$ one can construct
finite gauge theories (see next chapter).
Let us stress that the $\beta$-function of the magnetic dual of a 
finite electric theory
is not universally determined (i.e. we do not know a priori 
which value of $\mu_{\rm matter}$
gets mapped to a finite magnetic theory). Also, the self-dual 
point is non-universal.
As discussed before, SQCD with $G=SU(N_c)$ and $N_f$ matter fields in the
fundamental plus antifundamental representations has 
$\mu_{\rm gauge}=N_c$ and $\mu_{\rm matter}
=N_f$. 
Adding chiral matter fields in the adjont representation 
of $SU(N_c)$, they contribute with
$\mu_{\rm matter}({\rm adj})=N_c$ to the $\beta$-function; chiral fields in
the symmetric tensor representation of dimension ${1\over 2}N_c(N_c+1)$ 
contribute with
$\mu_{\rm matter}({\rm sym})={N_c\over 2}+1$, 
whereas antisymmetric tensor field matter
fields contribute with $\mu_{\rm matter}({\rm antisym})={N_c\over 2}-1$.
For $G=SO(10)$
we have that $\mu_{\rm gauge}=8$, $\mu_{\rm matter}({\rm vector})=1$
and $\mu_{\rm matter}({\rm spinor})=2$.

\section{Finite $N=1$ Gauge Theories}

There exist two known claims on the way how all loop
finiteness can be achieved, which we shall address in the present chapter.

\subsection{All order finiteness theorem}

Let us consider a chiral, anomaly free,
$N=1$ globally supersymmetric
gauge theory based on a group G with gauge coupling
constant $g$. The
superpotential of the theory is given by
\bea
W&=& \frac{1}{2}\,m_{ij} \,\phi_{i}\,\phi_{j}+
\frac{1}{6}\,C_{ijk} \,\phi_{i}\,\phi_{j}\,\phi_{k}~,
\label{supot}
\end{eqnarray}
where $m_{ij}$ and $C_{ijk}$ are gauge invariant tensors and
the matter field $\phi_{i}$ transforms
according to the irreducible representation  $R_{i}$
of the gauge group $G$.

The $N=1$ non-renormalization theorem \cite{nonre} ensures that
there are no mass
and cubic-interaction-term infinities.
As a result the only surviving possible infinities are
the wave-function renormalization constants
$Z^{i}_{j}$, i.e.,  one infinity
for each field. The one-loop $\beta$-function of the gauge
coupling $g$ is given in eq.(\ref{betag}).
 The $\beta$-functions of
$C_{ijk}$,
by virtue of the non-renormalization theorem \cite{nonre},
are related to the
anomalous dimension matrix $\gamma^i_{j}$ of the matter fields
$\phi_{i}$ as:
\be
\label{betsup}
\beta_{ijk} =
 \frac{d C_{ijk}}{d t}~=~C_{ijl}\,\gamma^{l}_{k}+
 C_{ikl}\,\gamma^{l}_{j}+
 C_{jkl}\,\gamma^{l}_{i}~.
\label{betay}
\end{equation}
At one-loop level $\gamma^i_{j}$ are \cite{PW}
\be
\gamma_{j}^{(1)i}=\frac{1}{32\pi^2}\,[\,
C^{ikl}\,C_{jkl}-2\,g^2\,C_{2}(R_{i})\delta^i_{j}\,],
\label{gamay}
\end{equation}
where $C_{2}(R_{i})$ is the quadratic Casimir of the representation
$R_{i}$, and $C^{ijk}=C_{ijk}^{*}$.

Therefore, all the one-loop $\beta$-functions of the theory vanish
if $\beta_{g}^{(1)}$  and $\gamma_{ij}^{(1)}$ vanish,
i.e.
\begin{equation}
\sum _i \ell (R_i) = 3 C_2(G) \,,
\label{1st}
\end{equation}

\begin{equation}
C^{ikl} C_{jkl} = 2\delta ^i_j g^2  C_2(R_i)\,,
\label{2nd}
\end{equation}
A very interesting result is that the conditions (\ref{1st},\ref{2nd}) are
necessary and sufficient for finiteness at
the two-loop level \cite{PW}.


The one- and two-loop finiteness conditions (\ref{1st},\ref{2nd}) restrict
considerably the possible choices of the irreps. $R_i$ for a given
group $G$ as well as the Yukawa couplings in the superpotential
(\ref{supot}).  Note in particular that the finiteness conditions cannot be
applied to the supersymmetric standard model (SSM), since the presence
of a $U(1)$ gauge group is incompatible with the condition
(\ref{1st}), due to $C_2[U(1)]=0$.  This naturally leads to the
expectation that finiteness should be attained at the grand unified
level only, the SSM being just the corresponding, low-energy,
effective theory.


A natural question to ask is what happens at higher loop orders.  The
answer is contained in a theorem \cite{LPS} which states the necessary
and sufficient conditions to achieve finiteness at all orders.  Before
we discuss the theorem let us make some introductory remarks.  The
finiteness conditions impose relations between gauge and Yukawa
couplings.  To require such relations which render the couplings
mutually dependent at a given renormalization point is trivial.  What
is not trivial is to guarantee that relations leading to a reduction
of the couplings hold at any renormalization point.\footnote{For 
a discussion  of dualities in
connection with the reduction of couplings see \cite{oehme1}.}
The necessary,
but also sufficient, condition for this to happen is to
require that such relations are solutions to the RE's
\be
\beta_g {d C_{ijk}\over dg} = \beta_{ijk}
\label{redeq2}
\end{equation}
and hold at all orders \cite{Zimmer}. Remarkably the existence of
all-order solutions to (\ref{redeq2}) can be decided at the one-loop
level \cite {Zimmer}.

Let us now turn to the all-order finiteness theorem \cite{LPS}, which
states under which circumstances a $N=1$ supersymmetric gauge theory 
can become finite to
all orders in the sense of vanishing $\beta$-functions, that is of
physical scale invariance.  It
is based on (a) the structure of the supercurrent in $N=1$ 
supersymmetric gauge theories \cite{fz-npb87,
pisi-npb196,pisi-book}, and on
(b) the non-renormalization properties of $N=1$ chiral anomalies
\cite{LPS,pisi}.
Details on the proof can be found in refs. \cite{LPS} and further
discussion in refs.~\cite{pisi,LZ,piguet}.  Here, 
we will just discuss the main points of this theorem.

One-loop finiteness, i.e. vanishing of the $\beta$-functions at one-loop,
implies that the Yukawa couplings $C_{ijk}$ must be functions of
the gauge coupling $g$. To find a similar condition to all orders it
is necessary and sufficient for the Yukawa couplings to be a formal
power series in $g$, which is solution of the RGE's (\ref{redeq2}).

We can now state the theorem for all-order vanishing
$\beta$-functions.
\bigskip

\noindent{\bf Theorem:}

Consider an $N=1$ supersymmetric Yang-Mills theory, with simple gauge
group. If the following conditions are satisfied

\begin{enumerate}
\item There is no gauge anomaly.
\item The gauge $\beta$-function vanishes at one-loop
  \be
  \beta^{(1)}_g = 0 =\sum_i l(R_{i})-3\,C_{2}(G).
  \ee
\item There exist solutions of the form
  \be
  C_{ijk}=\rho_{ijk}g,~\qquad \rho_{ijk}\in\complex
  \label{soltheo}
  \ee
to the  conditions of vanishing one-loop matter fields anomalous dimensions
  \be
  \gamma^{i~(1)}_j=0=\frac{1}{32\pi^2}~[ ~
  C^{ikl}\,C_{jkl}-2~g^2~C_{2}(R_{i})\delta_{ij} ] .
  \ee
\item these solutions are isolated and non-degenerate when considered
  as solutions of vanishing one-loop Yukawa $\beta$-functions:
   \be
   \beta_{ijk}=0.
   \ee
\end{enumerate}

Then, each of the solutions (\ref{soltheo}) can be uniquely extended
to a formal power series in $g$, and the associated supersymmetric
gauge theories
models depend on the single coupling constant $g$ with a $\beta$
function which vanishes at all-orders.
\bigskip

It is important to note a few things:
The requirement of isolated and non-degenerate
solutions guarantees the
existence of a formal power series solution to the reduction
equations (\ref{redeq2}).
The vanishing of the gauge $\beta$-function at one-loop,
$\beta_g^{(1)}$, is equivalent to the
vanishing of the R current anomaly.  The vanishing of
the anomalous
dimensions at one-loop implies the vanishing of the Yukawa couplings
$\beta$-functions at that order.  It also implies the vanishing of the
chiral anomaly coefficients.  This last property is a necessary
condition for having $\beta$ functions vanishing at all orders.




\subsection{The exact $\beta$-functions }

 The  second method of searching for all loop finite theories was
suggested some time ago in \cite{ermushev} and was reviewed recently in
\cite{leigh}. It is based on the all orders relation among the
gauge $\beta$-function $\beta_g$ and the anomalous dimensions of the
superfields $\gamma$ which was first derived using instanton
calculus \cite{NSVZ} given by
\be
\label{exact}
\beta_g^{NSVZ} \propto [{\sum_i l(R_i) - 3 C_2(G)} - \sum_i
l(R_i) \gamma_i],\label{gaugebnsvz}
\ee
which is claimed also to hold non-perturbatively 
\cite{leigh}. In addition the
second method is based on the general relation among the $\beta$-functions
for the Yukawa couplings and anomalous dimensions of the superfields given
in (\ref{betsup}) in accordance with the non-renormalization
theorem \cite{grisaru}.

Now observe that both $\beta$-functions,
the gauge $\beta$-function eq.(\ref{gaugebnsvz})
and the Yukawa $\beta$-function eq.(\ref{betay}), are linear functionals of
the anomalous dimensions of the matter superfields. These anomalous
dimensions now are complicated functions of the couplings, however the
relations among $\beta$'s and $\gamma$'s are very simple.

The criterion for having a finite theory is that all $\beta$-functions 
eqs.(\ref{gaugebnsvz}) and (\ref{betay})
vanish simultaneously. This puts $n$ constraints on the $n$ couplings
$g$, $C_{ijk}$. In the case that these constraints are linearly independent,
then one expects that their solutions are isolated points in the space of
couplings.
However if the one-loop gauge $\beta$-function vanishes, i.e. $\sum_il(R_i)-3C_2(G)=0$,
and
if only $p$ constraints are linearly
independent, then one is led to an $n - p$ dimensional manifold of fixed
points. Thus, if some of the $\beta$-functions are linearly dependent, and
there is at least one generic fixed point somewhere in the space of
couplings, then the theory will have a manifold of fixed points.
The theories
having couplings chosen to lie
on this manifold of fixed
points are interacting and finite in the sense
that all the $\beta$ functions, which are the physically relevant
quantities, vanish.
Of course
it is always possible that the constraints have no solutions at all,
as for instance when they put contradictory conditions on the anomalous
dimensions.

 In practice, when searching for all loop finite theories the second method
gives a faster answer, since it does not require calculation of the the
anomalous dimensions even at one-loop. However in general one has to be
careful. The first method is certainly better for real calculations in a
given theory and gives unambiquous results, but requires more work.

 Before closing our reference to the second method it is worth mentioning
the relation among the $\beta_g^{NSVZ}$ and the $\beta_g$ and $\gamma$, when
calculated in the DRED scheme. The relationship between $\beta_g^{NSVZ}$
and $\beta_g^{DRED}$ has been explored recently \cite{jackjones}, with
the conclusion that there exist an analytic redefinition of $g$, $g 
\longrightarrow
g'(g,C)$ which connects them, however the two schemes start to deviate at
three loops.

\section{Duals of $SU(N)$ finite models with vectors and tensors}


 In the first examples on duality transformations in four dimensional $N=1$
supersymmetric gauge theories, presented in
\cite{seiberg,seibergso,
intrisp},
the dual pairs were gauge theories based on the classical groups
$SU(N)$, $SO(N)$, $Sp(N)$ containing a single group factor and matter fields in
the fundamental plus anti-funtamental irreps for $SU(N)$, vector for $SO(N)$
and fundamental for $Sp(N)$, i.e. with matter fields in the defining
representation of the corresponding gauge group.
Among the possible generalizations of these constructions interesting dual
pairs have been found containing tensor matter fields
\cite{TensorKutasov},\cite{sakai},
paving the way to search for duals of realistic GUTs. A quite complete
list of dual pairs has been presented in \cite{Brodie+Strassler} containing
matter fields in two-index tensor representations in addition to
the fields in the defining representation of the above gauge groups.
A common feature of these constructions is the fact that they relate pairs
of theories of similar type. In addition they all contain gauge singlet
mesons in the dual superpotentials and in all the self-dual points
there exist marginal operators, which take the form of meson mass terms.

 The above classification of dual pairs allows us to examine all one loop
finite chiral models based on the $SU(N)$ gauge group with matter fields in
one conjugate symmetric, one antisymmetric and one adjoint representations
in addition to fields transforming according to the fundamental or
anti-fundamental ones in the prospect of making them all loop finite. In
general, to do so we have to add in the superpotential, which in
\cite{Brodie+Strassler} is
restricted in gauge invariant renormalizable terms made out of the tensor
fields also the corresponding ones involving combination of tensors and
fundamentals.
 In the following, first we 
briefly recall SQCD with only fundamental matter fields. Then we  
 apply the methods of
\cite{Brodie+Strassler} in a vector-like one
loop finite $SU(N)$ model with two fields in the adjoint representation, and
third we present representative examples of chiral models as well as the
basic features of the general $SU(N)$ models with the above particle
content. Any chiral model of this class, listed in
\cite{HPS}, can be examined in
a straightforward way following our examples.

\subsection{A simple example}

Let us first discuss the application of the method described in section 3.2
to a simple example, namely SQCD.
SQCD has played a major role in the recent developments
of duality and other exact results in $N=1$ supersymmetric
gauge theories, as reviewed in section 2.
Since we are looking for a model which allows for gauge invariant
Yukawa couplings, so we are restricted to $N_c=3$ where the
baryons are cubic and can be added to the superpotential.
Then the condition that the one-loop $\beta$ function vanishes
yields $N_f=9$. This model was analysed in
\cite{leigh}. Let us sumarize the results.

The model under consideration has the superpotential
\be
W=h\left(
\fQn{1}\fQn{2}\fQn{3}+\fQn{4}\fQn{5}\fQn{6}+\fQn{7}\fQn{8}\fQn{9}+
\aQn{1}\aQn{2}\aQn{3}+\aQn{4}\aQn{5}\aQn{6}+\aQn{7}\aQn{8}\aQn{9}\right)
\ .
\ee
One can use the exact $\beta$-function (\ref{exact}) to show that this is
finite.
First note that the above superpotential preserves a global
$[SU(3)^3\semidirect S_3]^2\semidirect Z_2$ subgroup of
the $[SU(9)]^2\semidirect Z_2$ flavour symmetry. Since the
$Q$ and $\bar{Q}$ still form irreducible representations of 
this global symmetry group, it is ensured that they all have the
same anomalous dimension.
Recalling the exact formulae for the gauge and
Yukawa $\beta$-functions
(\ref{exact}), (\ref{betsup}) we obtain that
in this case 
\be 
\label{gau}
\beta_{gauge} =\beta_g\propto 3/2 \gamma
\ee
\be
\label{YUK}
\beta_{Yuk} =\beta_h\propto 3/2 \gamma.
\ee
The eqs.(\ref{gau}), (\ref{YUK}) 
are obviously linearly dependent and hence according
to the discussion of \cite{leigh} reviewed in 3.2  the theory is finite,
for some relation among the couplings $h$ and $g$. To actually
find the required relation among $h$ and $g$ one would have
to calculate
the $\gamma$ order by order.

The dual is an $SU(6)$ gauge
theory with 9 flavors, 81 meson singlets $M$ and superpotential
\be
\tilde{W}=M^r_s  q_r\ot q^s + 
( B^{123}+B^{456}+B^{789}+\ot B_{123}+\ot B_{456}+\ot B_{789})
\ee
where $B^{123}\propto q_4q_5q_6q_7q_8q_9$.
This is an asymptotically free  theory
and hence strongly coupled in the infrared.
In the infrared the original finite theory and its asymptotically
free dual describe the same physics. Note that by inclusion
of the superpotential the formerly border case
of $N_f=3 N_c$ now has interesting infrared physics, too.

We see that the dual theory of a finite gauge theory is
not necesseraly finite.
Nevertheless it
would be much more interesting to find a dual which is also finite.
In this case one would expect the usual $N=1$ duality to 
be a strong-weak coupling S-duality valid at all scales.

\subsection{A vector-like finite $SU(N_c)$ gauge theory with two adjoints}

Let us  consider a vector-like $N=1$ theory, based on the gauge group
$SU(N_c)$ with two fields in the adjoint representation. For this model to
have vanishing gauge $\beta$-function we have to add equal number of left-
and right-handed matter fields in the fundamental representation and
moreover, according to (\ref{betag}) the number of flavours, $N_f$ 
should be equal to
the number of colours, $N_c$. The model with this superfield content has
symmetry group
\begin{eqnarray}
G= SU(N_c)_{local} \times
 [SU(N_c)_L \times SU(N_c)_R \times U(1)_B \times U(1)_R]_{global}
\end{eqnarray}
under which the superfields transform as
\begin{eqnarray}
Q &\sim& \bigl( N_c ; N_c,1;1,1/2 \bigr), \\
\bar{Q} &\sim& \bigl( \bar{N_c};1,\bar{N_c};- 1,1/2 \bigr), \\
X &\sim& \bigl( N_c^2-1;1,1;0,1 \bigr), \\
Y &\sim& \bigl( N_c^2-1;1,1;0,1/2 \bigr). 
\end{eqnarray}

The tree level superpotential is
\be
 W= Tr X^3 /3 + Tr X Y^2 + 
 f_1 \bar{Q}
X Q + f_2 \bar{Q} Y Q.
\ee
where the global symmetries are explicitly broken for non-vanishing
$f_1$, $f_2$.

The theory with the above superpotential, having vanishing one-loop
beta-function can become all orders finite requiring first that the
anomalous dimensions of all fields vanish and then checking if the
uniqueness requirement of the theorem of section 3.1 is satisfied.
We find that
the theory is finite to all orders only if $f_1=g^2$ and $f_2=0$.
 The dual of this model is based on the symmetry group
\be
 \tilde{G}= SU(5N_c)_{local}
 \times [SU(N_f)_L \times SU(N_f)_R \times U(1)_B \times U(1)_R]_{global}
\ee
The matter superfield content of the dual theory transforms as follows
\begin{eqnarray}
 q &\sim& (5N_c;\bar{N_c},1;1/2,0), \\
 \bar{q} &\sim& ({\overline{5N_c}};1,N_c;-1/2,0), \\
 X' &\sim& (25N_c^2 - 1;1,1;0,1), \\
 Y' &\sim& (25N_c^2 - 1;1,1;0,1/2). 
\end{eqnarray}
The dual theory with the above particle content is asymptotically free.
The dual superpotential has the form
\begin{eqnarray}
 \tilde{W}&= Tr X'^3 /3+Tr X'  Y'^2  +
 M_0 q X' Y'^2 \bar{q}+
M_1 q X' Y' \bar{q} +M_2 q X' \bar{q}
\end{eqnarray}
$$+N_0q Y'^2 \bar{q} + N_1 qY' \bar{q} +N_2 q \bar{q} +f_1 N_0 +f_2 M_1
$$
where $M_0$, $M_1$ and $M_2$ are singlet fields in the dual theory and
correspond to the operators $Q \bar{Q}$, $Q Y \bar{Q}$ and
$Q Y^2 \bar{Q}$ of the original theory respectively.
In the same way $Q X \bar{Q}$, $Q X Y \bar{Q}$ and $Q X Y^2 \bar{Q}$
get mapped to $N_0$, $N_1$ and $N_2$.
For the finite theory $f_1 \neq 0$, $f_2 = 0$, 
the dual gauge theory is broken to $SU(2N_c)$. The
resulting theory has 2 adjoints and after Higgsing a total
of $8 N_c$ fundamentals and antifundamentals.
Now the theory with the above particle content is asymptotically non-free.
 Note that the dual theory is not finite.

\subsection{Chiral finite $SU(N_c)$ gauge theories}

 Next let us consider chiral theories finite theories based on the $SU(N)$
gauge group and containing matter fields in the adjoint, antisymmetric and
conjugate symmetric tensors in addition to those in the fundamental and
anti-fundamental representations.
 Specifically consider an $SU(N_c)$, $N_c>3$ gauge theory with the following
symmetry group
\be
 G= SU(N_c)_{local} \times [SU(N_f)_L \times
  SU({N_f'})_R \times  U(1)_Y \times  U(1)_B \times  U(1)_R]_{global}
\ee
and particle content transforming according to

\begin{eqnarray}
 Q &\sim& ( N_c;N_f,1;\frac{6}{N_f}-1, \frac{1}{N_c}, 
1-\frac{N_c + 6k}{N_f (k+1)}), \\
\bar{Q} &\sim& ( \bar{N_c};1;\bar{N}_f'; \frac{6}{{N}_f'}-1,
- \frac{1}{N_c},
1-\frac{N_c - 6k}{{N}_f' (k+1)}     ), \\
 X &\sim& ( N_c^2 - 1;1,1;  0,0,\frac{2}{k+1}           ), \\
 Y &\sim& ( N_c (N_c - 1)/2 ;1,1 ;  1,\frac{2}{N_c},\frac{2}{k+1}  ), \\
{Y}' &\sim& (\overline{ N_c (N_c + 1)/2};1,1;  -1,-\frac{2}{N_c},\frac{2}{k+1} ).
\end{eqnarray}
The theory is chiral and for $N_f = N_c + 4$ and ${N}_f' = N_c - 4$ the
theory has vanishing one-loop $\beta$-function. The superpotential is
\be
 W =  Tr X^{(k+1)}/(k+1) + Tr X Y Y' 
   + f_1\bar{Q} X Q + f_2 \bar{Q}\bar{Q} Y + f_3 Q Q{Y}'.
\ee
Again the full global symmetries are explicitely broken for nonzero 
$f_1$, $f_2$, $f_3$.
In order the theory to be renormalizable k has to be 1 or 2.
Then the theory 
with the above superpotential can become all loop finite in the
usual way.

 The dual of this theory is an $SU(3k(N_f+{N}_f')/2 - N_c)$ gauge 
theory with the following full symmetry
\be
\tilde{G} = SU(\tilde{N_c})_{local} 
\times [SU(N_f) \times SU({N_f'}) \times  
U(1)_Y \times U(1)_B \times U(1)_R]_{global} 
\ee
and particle content transforming according to

\begin{eqnarray}
 q &\sim& (\tilde{N_c};{N_f},1; -\frac{6}{N_f}+1, \frac{1}{\tilde{N}_c},
1-\frac{\tilde{N}_c + 6k}{N_f (k+1)}
 ), \\
\bar{q} &\sim& (\bar{\tilde{N_c}};1,\bar N_f';    
-\frac{6}{{N_f'}}-1,- \frac{1}{\tilde{N}_c},
1-\frac{\tilde{N}_c - 6k}{{N_f'} (k+1)}           ), \\
 \tilde X &\sim& (\tilde{N_c}^2 - 1;1,1; 0,0,\frac{2}{k+1}           ), \\
 \tilde Y &\sim& (\tilde{N_c}(\tilde{N_c} - 1)/2;1,1; -1, \frac{2}{\tilde{N}_c},
\frac{2}{k+1}       ), \\
\tilde Y' &\sim&(\overline{ \tilde{N_c}(\tilde{N_c} + 1)/2};1,1;1,1; 1,
-\frac{2}{\tilde{N}_c},
\frac{2}{k+1}     ).
\end{eqnarray}
In addition there are the meson fields $N$, $P$, $\tilde{P}$
and $M$.

The dual theory with the above gauge group and particle content is
asymptotically free and has the following superpotential
\be
\label{supk1}
 \tilde{W} = Tr \tilde{X}^{(k+1)}/(k+1) + Tr \tilde X \tilde Y\tilde{Y'}
    +
\ee
$$
+ \sum_{j=0}^{k-1} \left \{ N_j \bar{q} \tilde{X}^{k-j-1} q +
P_j q \tilde X^{k-j-1} \tilde{Y}' + \tilde{P} \bar{q} \tilde Y
\tilde X^{k-j-1}\bar{q}+
M_j \bar{q}\tilde X^{k-j-1} \tilde Y \tilde{Y}' q \right \}.
$$
$$ + f_1 M_1 + f_2 \tilde{P}_0 + f_3 P_0 $$

Let's examine the theory when $k=2$. In that case the dual
theory is an $SU(5N_c)$ gauge theory and when all $f_i$, $i=1,2,3$ are
non-vanishing the theory can break to $SU(3N_c - 2)$ at most. In the latter
case the theory contains $4N_c+8$ superfields in the fundamental
representation, $4N_c$ in the anti-fundamental, one superfield in the
adjoint and one in the conjugate symmetric representations, while there is
no remaining superfield in the antisymmetric rep. With this particle
content the theory is asymptotically non-free.

 Next let us discuss a couple of specific models based on $SU(N_c)$, one with
$N_c$ odd and the other with $N_c$ even which, as we shall see, distinquishes
further these models. All other models can then be examined similarly.

 Let us consider an $SU(5)$ gauge theory containing the chiral
supermultiplets \newline ${\bf ( 5, \bar{5}, 24, 10, \bar{15})}$ with
multiplicities ( 9, 1, 1, 1, 1 ). The model with this particle content has
vanishing one-loop $\beta$-function. The superpotential for $k=1$ becomes
\be
 W = (Tr 24^3)/3 + Tr 24 \; 10  \; \bar{15}
     + \sum_{i=1}^5 f_1 \bar{5}_i \; 24 \; 5_i +
 \sum_i f_{2,i} \bar{5}_i \;  
\bar{5}_i \; 10
     + \sum_i f_{3,i} 5_i \; 5_i \; \bar{15} 
\ee
This model can become all loop finite. Its dual is an $SU(25)$ gauge
theory, which contains the chiral supermultiplets ${\bf( 25, \overline{25}, 
224, 105, 120)}$ with multiplicities ( 9, 1, 1, 1, 1 ). The dual theory then 
looks from first sight rather unatractive. The superpotential
can be read from eq.(\ref{supk1}) and from the matching
of the gauge invariant operators by straightforward substitution of the 
corresponding fields.
The theory becomes simpler after considering its spontaneous
breakdown. Then the vevs $ <\overline{25}_i \; 224 \; 25_i>$,
which are enforced by the equations of motion from the
dual superpotential, 
break the
gauge group of the dual theory by one unit each, the vevs $<\overline{25}_i 
\; 105 \;
\overline{25}_i>$
break the gauge group by two units all together (since the
antisymmetric tensor field is absorbed via the Higgs mechanism) and the
vevs $<25_i  \; 120 \;
 25_i>$ break the gauge group by one unit each \cite{Li}.
Therefore the dual theory eventually is an $SU(13)$, which contains
one adjoint, one antisymmetric and one conjugate symmetric tensors and 20
superfields in the fundamental and 28 in the antifundamental reps. The
theory is asymptotically non-free.

 As a second concrete example let us consider an $SU(4)$ gauge theory. The
theory with particle content in the representations ${\bf ( 4, 15, 6 , 
\overline{10})}$
with multiplicities (8, 1, 1, 1) has vanishing one loop $\beta$-function.
A similar analysis as was done in the previous example is straithforward.
What is more interesting here is that the model belongs to the class of
models with $N_c = k n$. Then with $k = 2$  the theory has flat
directions
\cite{Brodie+Strassler},
 which break the group to $SU(2) \times SU(2)$. The dual gauge
group breaks from $SU(3k(N_f+{N_f'})/2 - N_c)$ down to
$SU(3(N_f+{N_f'}/2 - 2)^k$, i.e. in the present case to 
$SU(8) \times SU(8)$.

\section{Duals of Chiral Finite Models: Vectors and Spinors in SO(10)}

\subsection{The method and an instructive example}

This method of searching for duals applies to $N=1$ supersymmetric gauge
theories with gauge group $SO(10)$ and matter fields in $N_q$ spinor
representations and $N_f$ vector representations \cite{so10}. 
Its dual counterpart is a
chiral model with semisimple gauge group $SU(N_f +2N_q-7) \times 
Sp(2N_q-2)$.

It is instructive to consider first a relatively simple example
consisting of a $N=1$ supersymmetric $SO(10)$ model with two spinors and $N_f$
vector matter fields. The model has symmetry group:
\begin{eqnarray}
G=SO(10)_{local} \times [SU(N_f) \times SU(2) \times U(1)_Y
 \times U(1)_R]_{global}
\end{eqnarray}
under which the chiral superfields transform as
\begin{eqnarray}
V^i_\mu & \sim& \bigl( 10; \fund ,1 ; -4, \Rtwo \bigr) \\
Q^\A_\I  & \sim& \bigl( 16; 1,2 ; \Nf, \Rtwo \bigr) 
\end{eqnarray}
and has vanishing superpotential. With the above hypercharge and R-charge
assignements for the matter fields the model is free of global anomalies
and is is asymptotically free for $N_f < 20$.

Let us discuss first its confining phase. Assuming the following
hierarchy of successive spontaneous symmetry breakings
\be
{SO(10) \> {\buildrel 2(16) \over \longrightarrow} \>
  G_2 \> {\buildrel 10 \over \longrightarrow} \>
  SU(3) \> {\buildrel 10 \over \longrightarrow} \>
  SU(2) \> {\buildrel 10 \over \longrightarrow} \>
1,} 
\ee
one can easily count the gauge invariant operators needed to act as
moduli space coordinates for small numbers of vector flavours. These gauge
invariant operators appear in the effective low energy description of the
theory. In table (\ref{parton})
the number of partonic degrees of freedom  and the
generic unbroken gauge subgroup 
$H_{local}$ as a function of $N_f$ are given. The
dimension of the coset space $G_{local}/H_{local}$ coincides with the number of
matter fields eaten by the Higgs mechanism. The number of remaining partons
listed in the table equals the number of independent hadrons which label
flat directions in the effective theory.
\begin{table}[h]
\label{parton}
\begin{center}
\begin{tabular}{|c|c|c|c|c|}
\hline
 $\Nf$ & Parton DOF & Unbroken Subgroup & Eaten DOF&
 Hadrons \\
 \hline
0 & 32 & $G_2$   & $45-14=31$ & 1 \\
1 & 42 & $SU(3)$ & $45-8=37$ & 5 \\
2 & 52 & $SU(2)$ & $45-3=42$ & 10 \\
3 & 62 & 1 & 45 & 17 \\
4 & 72 & 1 & 45 & 27 \\
5 & 82 & 1 & 45 & 37 \\
\hline
\end{tabular}
\end{center}
\caption{Number of gauge invariant operators which are the coordinates
of moduli space}
\end{table}

In order to construct expicitly the hadron fields we recall the tensor
product
\be
16 \times 16 = 10_\S + 120_\A + 126_\S
  = [1]_\S + [3]_\A + \widetilde{[5]}_\S.
\ee
    where $[n]$ denotes an antisymmetric rank-n tensor, the `S' and `A'
 subscripts indicate symmetry and antisymmetry under spinor exchange and
   the tilde over the last term implies that the rank-5 irrep is complex
  self-dual.
  One can then produce the following gauge invariant composites

\begin{eqnarray}
K &=& Q^\T_\I (\s_\xrm \s_2)_{\I\J} \Gamma^\mu C Q_\J
  Q^\T_\K (\s_\xrm \s_2)_{\K\L} \Gamma_\mu C Q_\L
  \sim \bigl(1; 1, 1; 4\Nf, 4 \Rtwo \bigr), \\
M^{(ij)} &=& (V^\T)^{i \mu} V^j_\mu
        \sim \bigl(1; \sym, 1; -8, 2 \Rtwo \bigr), \\
N^i_\xrm &=& Q^\T_\I (\s_\xrm \s_2)_{\I\J} \Vslash^i C Q_\J
        \sim \bigl(1; \fund,3; 2 \Nf - 4, 3 \Rtwo \bigr), \\
P^{[ijk]} &=& {1 \over 3!} Q^\T_\I (\s_2)_{\I\J} \Vslash^{[i} \Vslash^j
  \Vslash^{k]} C Q_\J
  \sim \bigl(1; \antithree,1; 2\Nf - 12, 5 \Rtwo \bigr), \\
R^{[ijkl]} &=& {1 \over 4!} Q^\T_\I (\s_\xrm \s_2)_{\I\J} \Gamma^\mu C Q_\J
        Q^\T_\K (\s_\xrm \s_2)_{\K\L} \Gamma_\mu \Vslash^{[i}
  \Vslash^j \Vslash^k \Vslash^{l]} C Q_\L \\
&\sim& \bigl(1; \antifour, 1; 4\Nf - 16, 8 \Rtwo \bigr),\\
T^{[ijklm]}_\xrm &=& {1 \over 5!} Q^\T_\I (\s_\xrm \s_2)_{\I\J}
  \Vslash^{[i} \Vslash^j \Vslash^k \Vslash^l \Vslash^{m]} C Q_\J \\
        &\sim& \bigl(1; \antifive,3; 2\Nf - 20, 7 \Rtwo \bigr), \\
\end{eqnarray}
where Greek, small Latin and large Latin letters respectively denote 
$SO(10)$, $SU(N_f)$ and $SU(2)$ spinor indices.

Comparing the number of composite operators with the number of
independent flat directions as a function of $N_f$,
one finds that $K$,$M$,$N$
and
P account for all massless fields in the $SO(10)$ model up to three
flavours. For $N_f$ larger or equal to 4 an increasing number of constraints
is also needed, which can be imposed in the superpotential form using
Lagrange multipliers.

Turning to the search of the dual counterpart of this model we should
recall that there exist a class of dual pairs
\cite{pouliot} in which one member is a
$N=1$ supersymmetric model with gauge group $G_2$ with $N_f$ matter fields in
the fundamental representation and its dual is based on $SU(N_f-3)$ gauge
group. Since the $SO(10)$ model with two spinors reduces to the above $G_2$
model along a flat direction, where both spinors acquire vev, it is
natural to start searching for its dual by looking for extensions of the
$SU(N_f-3)$, which is dual to $G_2$. After exploring several possibilities the
authors in \cite{so10}
concluded that the dual of $SO(10)$ is not a simple group. The
simplest generalization found was that the dual has symmetry group
\be
  \tilde{G}=[SU(N_f-3) \times Sp(2)]_{local} \times
[SU(N_f) \times SU(2) \times U(1)_Y \times U(1)_R]_{global} 
\ee
  with superfield content
\begin{eqnarray}
q^\a_i &\sim& \bigl( \fund ,1 ; \antifund,1; 2{\Nf-6 \over \Nf-3} ,
  {5 (\Nf-4) \over (\Nf-3)(\Nf+4)} \bigr) \\
\qp^{\a\adot}_\I &\sim& \bigl( \fund,2; 1,2; -{2 \Nf \over \Nf-3},
  {(\Nf+2)(\Nf-4) \over (\Nf-3)(\Nf+4)} \bigr) \\
\qbar^\xrm_\a &\sim& \bigl( \antifund,1; 1,3; -2\Nf{\Nf-4 \over \Nf-3},
  - {\Nf^2-18 \Nf + 40 \over (\Nf-3)(\Nf+4)} \bigr) \\
s_{\a\b} &\sim& \bigl( \symbar,1; 1,1; {4\Nf  \over \Nf-3},
  2 {3 \Nf-4  \over (\Nf-3)(\Nf+4)} \bigr) \\
t^{\adot}_\I &\sim& \bigl( 1,2; 1,2; 2\Nf, 2 \Rtwo \bigr) \\
m^{(ij)} &\sim& \bigl(1,1; \sym,1; -8, 2 \Rtwo \bigr) \\
n^i_\xrm  &\sim& \bigl(1,1; \fund,3; 2\Nf-4, 3 \Rtwo \bigr) \\
\end{eqnarray}
and tree level superpotential
\be
\tilde{W}={1 \over \mu_1^2} m^{(ij)} q^\a_i s_{\a\b} q^\b_j
 + {1 \over \mu_2^2} n^i_\xrm  q^\a_i \qbar^\xrm_\a
 + \lambda_1 \e_{\adot \bdot} \e^{\I\J} {\qp}^{\a\adot}_\I s_{\a\b}
    {\qp}^{\b\bdot}_\J
 + \lambda_2 \e_{\adot \bdot} {\qp}^{\a\adot}_\I (\s_\xrm \s_2)^{\I\J}
  \qbar^\xrm_\a t^{\bdot}_\J.
\ee
Then the above construction was generalized to $N=1$ supersymmetric $SO(10)$
gauge theory with arbitrary $N_q$ spinor representations and $N_f$ vector
representations.

\subsection{The dual of a finite $SO(10)$ gauge theory}

 Next let us apply the generalized method to an $SO(10)$ finite model.
Consider a $N=1$ supersymmetric model based on the gauge group $SO(10)$ with
matter fields in $N_f=8$ vector and $N_q=8$ spinor representations. The model
with this superfield content has vanishing one-loop gauge $\beta$ function
and symmetry group
\be 
 G=SO(10)_{local} \times [SU(8) \times SU(8) \times
U(1)_Y \times U(1)_R]_{global} 
\ee 
under which the superfields transform as
 
\begin{eqnarray}
 V_{i,\mu} &\sim& (10;8,1;-16,2/3) \\
 Q_{A,I} &\sim&  (16;1,8;8,2/3).
\end{eqnarray}
Then its dual is  found to be based on the gauge symmetry group
\be 
 \tilde{G} =[SU(17) \times Sp(14)]_{local} \times [SU(8) \times
 SU(2) \times U(1)_Y \times U(1)_R]_{global} 
\ee
It is worth noting that only an $SU(2)$ subgroup of the global $SU(8)$ that
rotates the spinors in the original theory is realized in the ultraviolet
of its dual. The $SU(2)$ subgroup is embedded inside $SU(8)$ so that the
fundamental 8-irrep of the latter is mapped on the 8-dimensional irrep of
the former. The full $SU(8)$ global symmetry is realized in the dual theory
only at long distances.

The matter content of the dual theory  has the following transformation
properties
 
\begin{eqnarray}
 q_{\alpha}^i &\sim& (17,1;\bar{8},1;160/17,-14/51), \\
 q_{\alpha,\adot}^{'I} &\sim& (17,14;1,2;-112/17,20/51),\label{e5116} \\
 \bar{q}^{\alpha}_{(I-1...I-2N-q-2)} &\sim&
 (\overline{17},1;1,15;-160/17,14/51), \\
 s_{\alpha,\beta} &\sim& (\overline{153},1;1,1;224/17,62/51),\label{e5118} \\
 t^{\adot}_{(I-1...I-13)} &\sim& (1,14;1,14;16,4/3), \label{e5119}\\
 m^{(ij)} &\sim& (1,1;\overline{36},1;-32,4/3), \\
 n^i_{(I-1...I-14)} &\sim& (1,1;8,15;0,2).
\end{eqnarray}
The tree level superpotential of the dual theory is
\be 
\tilde{W}=\frac{1}{\mu_1^2} mqsq+ \lambda_1 q' s q' + \frac{1}{\mu_2^2} 
q n \bar{q} +  \lambda_2 q' \bar{q} t 
\ee
The
coeficients $\mu_{1,2}$ and $\lambda_{1,2}$
represent dimensionful and dimensionless
couplings. The mesons $m$ and $n$ are gauge singlets
of the dual gauge group
correponding to the gauge invariant operators
$V V$ and $ V Q Q$ respectively of the original gauge group.

There are some remarks to be made concerning this dual. The first is that
it satisfies all necessary anomaly checks including the anomaly matching
conditions associated with the common $SU(8) \times SU(2) \times
 U(1)_Y \times U(1)_R$
global symmetry group. The second  concerns the  $\beta$ functions of the dual
product gauge groups. The first factor has $\beta$ function
coefficient $b^{(1) SU(17)}_0=16$ and therefore it is assymptotically
free, while the second vanishes i.e. is finite at one loop.
 
Now in order to make the original theory finite
to all orders we have to add the superpotential
\be
W=h \sum_{i=1}^8 Q_iQ_iV_i,
\ee
($i$ labeling the 8 flavors)
which is the simplest choice having a universal Yukawa coupling.
This
can easily be seen to be all-loop finite.
One way to see this is to apply the all-loop finiteness theorem
of 3.1. The 1-loop $\beta$ function for the Yukawa
coupling reads
\be
\beta^{(1)}_{h}=h (14 |h|^2 - \frac{63}{2} g^2)
\ee
and the 1-loop $\gamma$ functions are
\be
\gamma^{(1)}_{10}=\left [ 4 |h|^2 - 9 g^2 \right ]
\ee
and
\be
\gamma^{(1)}_{16} = \left [ 5 |h|^2-\frac{45}{4} g^2 \right ]
\ee
which obviously vanish simultanousely for
\be
|h|^2= \frac{9}{4} g^2
\ee
and satisfy the criteria of the theorem 3.1.

Finiteness can also be established according to the method
of \cite{leigh} by noting that the gauge and the
Yukawa $\beta$ functions
\begin{eqnarray}
\beta_{g}  &\propto& 8 \gamma_V + 16 \gamma_Q \\
\beta_h &\propto& \gamma_V +2 \gamma_Q \\
\end{eqnarray}
are linearly dependent.

Now let us study the finiteness on the dual side. 
The complete dual superpotential is
\be
\tilde{W}= \frac{1}{\mu_1^2} m qsq +  \lambda_1 
q'sq' + \frac{1}{\mu_2^2} n q \bar{q} + \lambda_2 q' \bar{q} t +
\tilde{h} n
\ee
The equations of motion for $N$ enforce $<q \bar{q}>$ to be
nonzero and of rank 8 since we included the sum over all 8 flavors in
the original theory. This breaks the dual theory to $SU(9)$ while
eating 8 $q$, $\bar{q}$ flavors via the Higgs mechanism. It is easy
to work out the matter content charged under the unbroken
gauge group. The remaining symmetry group is
\be
 \tilde{G} =[SU(9) \times Sp(14)]_{local} \times [SU(8) \times
 SU(2) \times U(1)_R \times U(1)_Y]_{global}
\ee
with the charged (under $SU(9) \times Sp(14)$) fields transforming like
\begin{eqnarray}
 q'_1 &\sim& (1,14;8,2;2/3,-16), \\
 q'_2 &\sim& (9,14;1,2;4/27,16/9), \\
 \bar{q} &\sim&
 (\overline{9},1;1,7;14/27,-160/9), \\
 s_1 &\sim& (\overline{45},1;1,1;46/27,-32/9), \\
 s_2 &\sim& (\overline{9},1;\bar{8},1;32/27,128/9), \\
 t_1&\sim& (1,14;1,6;4/3,16), \\
 t_2&\sim& (1,14;1,8;4/3,16), 
\end{eqnarray}
where $q'_1$ and $q'_2$ come from $q'$ in
eq.(\ref{e5116}), $s_1$ and $s_2$ from $s$ in eq.(\ref{e5118}) and finally
$t_1$ and $t_2$ from $t$ in eq.(\ref{e5119}).
They interact via a superpotential
\be
\tilde W= \lambda_1(
q'_2 s_1 q'_2  + q'_1 q'_2 s_2) + 
\lambda_2(q'_2\bar{q} t_1+\bar{q} q'_2 t_2)
\label{dualso10sup}
\ee
The resulting product gauge group now has vanishing one-loop
$\beta$ function for both gauge factors! This is very interesting,
since it seems to indicate that we this way created
a duality between two finite theories. 
In addition to the above charged
fields, there will be  in general also gauge singlet
fields. These gauge singlets will introduce new interactions in the
superpotential which are quite cumbersome to be worked out.
Note however that these gauge singlet fields would spoil finiteness. 
So we propose
that indeed the charged sector from above describes
the full dual theory and all the gauge singlets actually decouple.

The following facts support this
scenario:
\begin{itemize}
\item With the superpotential (\ref{dualso10sup})
the magnetic theory is all-loop finite. It is
easy to verify this with the approach of \cite{leigh}.
The gauge and Yukawa $\beta$ functions are
\bea
\beta_{SU(9)} &\sim& 28 \gamma_{q'} + 7 \gamma_{\bar{q}} + 19 \gamma_{s}, \\
\beta_{Sp(14)} &\sim& 34 \gamma_{q'} + 14 \gamma_{t} , \\
\beta_{\lambda_1} &\sim& 2 \gamma_{q'} +  \gamma_{s}, \\
\beta_{\lambda_2} &\sim&  \gamma_{q'}+\gamma_{\bar{q}} +\gamma_{t} 
. 
\eea
Since
$$ 2 \beta_{SU(9)} + \beta_{Sp(14)} -38 \beta_{\lambda_1} -14
\beta_{\lambda_2}  =0$$
 all $\beta$-functions  are linearly dependent, and
hence according to \cite{leigh} we got a finite theory
in the sense that all $\beta$-functions vanish.
\item The anomalies for $U(1)^3$, $U(1)$, $U(1)_R^3$, $U(1)_R^{} $,
 $U(1)_R^2 U(1)$, $U(1)_R^{} U(1)^2$ match those in the
original theory. This is a highly non-trivial check.\footnote{For the
non-abelian symmetries things are much more complicated.
Already in the unbroken theory of \cite{so10} the issue
of accidental symmetries \cite{acci} complicates the discussion
of the non-abelian global symmetries. 
Therefore we did not work out the anomaly matching with respect to the
non-abelian global symmetries.}


\end{itemize}

\section{Searching for duals of FUTs with the Deconfinement Method}

\subsection{The method and an instructive example}

This method of determining duals of given gauge theories can be used in
all theories containing superfields in tensor representations.

The basic
idea of the method \cite{berkooz} is to assume that the tensors
of the gauge theory under consideration are composite fields made out of
elementary fields transforming according to the fundamental representation
of another gauge group that confines and reproduces the tensorial
spectrum of the original theory. Thus the latter gauge group is
supposed to describe the fundamental theory and the former is only an
effective theory. Since it is known how to `deconfine' an arbitrary
two-index tensor, the method can be applied to all chiral models in the list of
\cite{HPS}, with the exception of $E_6$ models and $SO(10)$ with
spinors.

 The method has the advantage that it is straightforward, and the obvious
disadvantage that it introduces a new `fundamental' gauge group for
every tensor superfield present in the original theory. Therefore the dual
theories obtained by applying this method are usually rather complicated.

 In the following first we present a simple example following
\cite{pouasy}, in order to demonstrate how the method
works in practice, then we apply it in a simple chiral one-loop finite
$SU(3)$ model and then we find the dual of the realistic finite $SU(5)$ GUT.
The method can then applied in all other models of \cite{HPS} that contain
tensor superfields with the exception already mentioned.

 Let us first consider an $SU(N_c)$, $N = 1$ supersymmetric gauge theory with
the following matter superfields: $N_f$ in the fundamental, ${N_f'}$ 
in the anti-fundamental and one in the antisymmetric tensor representations.
The $SU(N_c)^3$ anomaly cancellation requires $N_f = {N_f'} - N_c + 4$.
The full symmetry group of the theory is then
\be
G = SU(N_c)_{local} \times [SU(N_f)_L \times SU({N_f'})_R
 \times U(1)_Y \times 
U(1)_B \times U(1)_R]_{global}
\ee
and the superfields are transforming according to
\begin {eqnarray}
 Q &\sim& ( N_c; N_f, 1 ; 1 , N_c - N_f , 2 - 6/N_c ),\\
\bar{Q} &\sim& (\bar{N_c} ; 1 ,{N_f'} ;-N_f/(N_c+N_f-4) ,
 m_f , 6/N_c ),\\
 Y &\sim& ( N_c(N_c+1)/2 ; 1 , 1 ; 0 , -2N_f, -12/N_c ).
\end{eqnarray}
To see how the method works consider a $N=1$ gauge theory based on the $
Sp((N_c - 3)/2)$ group ($N_c$ odd) containing $N_c + 1$ matter fields in the
fundamental representation say, $y_i$ and $z$ and $N_c$ singlets
$\bar{P_i}$, $i=1,...,N_c$. Therefore the theory has the following symmetry
\be
 G' = Sp((N_c - 3)/2)_{local}
 \times [ SU(N_f)_L \times SU(N_f)_R ]_{global}
\ee
and the matter fields transform  according to
\begin{eqnarray}
 y,z &\sim&( (N_c - 3)/2 ; N_c + 1 , 1 ),\\ 
\bar{P_i} &\sim& ( 1 ; 1 , N_c ).
\end{eqnarray}
This is a confining theory and provides a superpotential \cite{intrisp}
\be
 W = y^N z = A^{((N-1)/2)} P,
\ee
for the gauge invariant fields $A_{ij} = y_i y_j$ and $P_i = z y_i$.
Then one adds a mass term for $P_i$ and $\bar{P_i}$ in the superpotential 
in the form of a coupling $z y_i\bar{P_i}$, which breaks the global flavour 
symmetry
\be
 SU(N_c + 1) \times SU(N_c) \longrightarrow SU(N_c) \times U(1).
\ee
Integrating out the superfieds $P_i$, $\bar{P_i}$ one obtains a theory with
$N_c(N_c - 1)/2$ singlet superfields $A_{ij}$  without superpotential, i.e.
without constraints on the light superfields $A_{ij}$.

 Next consider gauging the $SU(N_c)$ flavour symmetry. Clearly under this
symmetry the superfield A transforms as an antisymmetric tensor. Then in
order to cancel the $SU(N_c)^3$ anomaly one has to introduce more
$Sp((N_c - 3)/2)$ singlets say, Q and $\bar{Q}$, which transform 
according to the fundamental and anti-fundamental reps of $SU(N_c)$.

 Therefore the above constructed $SU(N_c) \times Sp((N_c - 3)/2)$ theory is
equivalent to the original theory. The charge assignements in this expanded
theory can be determined from the superpotential and the definitions of
the composite fields and its full resulting symmetry is
\be
 G'' = [SU(N_c) \times Sp((N_c - 3)/2)]_{local} \times [SU(N_f)_L \times
SU(N_f)_R \times U(1)_Y \times U(1)_B \times U(1)_R]_{global}
\ee
The matter fields transform under the above symmetry group according to
\begin{eqnarray}
 y &\sim& ( N_c , (N_c - 3)/2 ; 1 , 1 ; 0 , -N_f , - 6/N_c ),\\
 z &\sim& ( 1 , (N_c - 3)/2  ; 1 , 1 ; 0 , N_f N_c , 8 ),\\
\bar{P} &\sim& (\bar{N_c} , 1 ; 1 ,1 ; 0 , N_f (1 - N_c) , -6 + 6/N_c ),\\
 Q &\sim& ( N_c , 1 ; N_f , 1 ; 1 , N_c - N_f , 2 - 6/N_c ),\\
\bar{Q} &\sim&
 (\bar{N_c} , 1 ; 1 , N_c + N_f - 4 ; - N_f/(N_c + N_f - 4) , N_f , 6/N_c ).
\end{eqnarray}
The $SU(N_c)$ gauge group is in a non-Abelian Coulomb phase for $N_f\geq 5$ ,
and  therefore can be dualized by the $SU(N_c)$ duality
prescription of \cite{seiberg}. The resulting dual model is a gauge 
theory based on
the 
\be
SU(N_f - 3) \times Sp((N_c - 3)/2) 
\ee
group with the following symmetry
\be
\tilde{G} = [SU(N_f - 3) \times Sp((N_c - 3)/2)]_{local} \times [SU(N_f)_L 
\times SU(N_f)_R]_{global}
\ee
with matter content transforming according to
\begin{eqnarray}
\tilde{x} &\sim& (\overline{(N_f - 3)} , (N_c - 3)/2 , 1 , 1 ),\\
 p &\sim& ( N_f - 3 , 1 ; 1 , 1),\\
\bar{q} &\sim& (\overline{(N_f - 3)} , 1 ;\bar{N}_f ,1 ),\\
 q &\sim& ( N_f - 3 , 1 ; 1 , (\bar{N}_c - N_f - 4)),\\
\bar{l} &\sim& ( 1 , (N_c - 3)/2 ; 1 , N_c + N_f - 4 ),
\end{eqnarray}
and
\begin{eqnarray}
 M = Q\bar{Q} &\sim& ( 1 ,1 ; N_f , N_c + N_f - 4 ),\\
 B_1 = Q A^{(N_c - 1)/2} &\sim& ( 1 , 1 ; N_f , 1),
\end{eqnarray}
where $M=Q\bar{Q}$ are mesons and $B_k = Q^k A^{((N-k)/2)}$ are
baryons.
The dual theory has superpotential
\be
\tilde{W} = M q\bar{q} + B_1 p\bar{q} +\bar{l}\tilde{x} q.
\ee
The gauge group $Sp((N_c - 3)/2)$ for $N_f \geq 5$ is in a non-abelian Coulomb 
phase and can be dualized by the $Sp$ duality according to \cite{intrisp}. The 
resulting dual theory after integrating out the massive
fields a gauge theory based on the group $SU(N_f - 3) \times Sp(N_f - 4)$ with 
full symmetry
\be
\tilde{G'} = [SU(N_f - 3) \times Sp(N_f - 4)]_{local}
 \times [SU(N_f)_L \times 
SU(N_f)_R \times U(1)_Y \times U(1)_B \times U(1)_R]_{global}
\ee
and matter fields transforming according to
\begin{eqnarray}
 x &\sim&(N_f - 3 , N_f - 4 ; 1 , 1 ; -N_f/(N_f - 3) , 0 ,- 1 )\\
 p &\sim&(N_f - 3 , 1; 1 , 1 ; -N_f/(N_f - 3) , N_c N_f , 6 )\\
\bar{\alpha} &\sim&(
\overline{(N_f - 3)(N_f - 4)/2} , 1 ; 1 , 1 ; 2N_f/(N_f - 3), 0 , 4 )\\
\bar{q} &\sim&(\overline{(N_f - 3)} , 1;\bar{N}_f , 1 ; 3/(N_f
 - 3) , -N_c , 0 )\\
 l &\sim&( 1 , N_f - 4 ; 1 ,\overline{(N_c + N_f - 4)} ; N_f/(N_c
 + N_f - 4) , 0 , 1 )\\
 M &\sim&( 1 ,1 ; N_f , N_c + N_f - 4 ; (N_c - 4)/(N_c + N_f - 4) , N_c , 2 )\\
 H &\sim&( 1, 1 ; 1 , (N_c + N_f -4)(N_c + N_f - 5)/2 ; -2N_f/(N_c
 + N_f - 4) , 0 , 
0 )\\
 B_1 &\sim&( 1 , 1 ; N_f , 1 ; 1 , N_c(1 - N_f) , -4 )
\end{eqnarray}
and superpotential
\be
\tilde{W'} = M\bar{q} l x + H l l + B_1 p\bar{q} +\bar{\alpha} x^2,
\ee
where
\be
 B_k = Q^k A^{((N-k)/2)} 
\ee
are baryons.
This is the most general superpotential allowed by the
symmetries, holomorphy and smoothness near the origin in field space.

 Armed with the above explicit construction we can first apply the method
to a chiral finite  model based on $SU(3)$ gauge group and then to a
realistic chiral finite SU(5) GUT.

\subsection{The dual of a finite chiral $SU(3)$}

One of the simplest chiral finite models that can be treated
with the deconfinement method is a $SU(3)$ gauge group with matter
transforming symmetric tensor representation and an appropriate 
number of fundamentals and anti-fundamentals to make it one-loop
finite and to cancel the anomaly, that is 1 symmetric tensor $S$,
3 fundamentals $Q$ and 10 antifundamentals
$\bar{Q}$. The deconfinement of theories
with symmetric 2-index tensors has been discussed in great detail
in \cite{sakai}.

First consider the theory without a tree-level superpotential.
We will add the superpotential that is necessary to make the theory
finite later on.
The symmetry group is
\be
G=SU(3)_{local} \times [SU(3) \times SU(10) \times 
U(1)_1 \times U(1)_2U \times (1)_R  ]_{global}
\ee
where the matter fields transform as:
\begin{eqnarray}
\label{electric;theory}
Q           &\sim&(3;3, 1,1,0,4), \\
\bar{Q}   &\sim  &(\bar{3}; 1,10,-3/10,3,0 ), \\
S           &\sim&(6; 1, 1,0,-6,0  ).
\end{eqnarray}

The symmetric tensor $S$ can be deconfined with the help of a $SO(8)$
theory. Now this $SO(8)$ has only fundamental matter and can be
dualized according to \cite{seibergso}. After dualizing
also the $SU(3)$ (which by the virtue of the deconfinement
has only fundamental matter), we obtain the following net dual with
symmetry group
\be
\tilde{G} = [ SU(8) \times SO(14) ]_{local} \times [SU(3) \times 
SU(10) \times U(1)_1  \times U(1)_2  \times U(1)_R]_{global} 
\ee
and matter content
\begin{eqnarray}
\label{dual2;theory}
\bar{q} &\sim&( \bar{8}; 1, \bar{3}, 1,-{5 \over 8}
,-3,-{5 \over 2} ), \\
p&\sim &(8; 1, 1, 1,-{3 \over 8},9,-{7 \over 2} ), \\
u &\sim& (1; 1, 1, 1,0,-18,6 ), \\
M &\sim& (1; 1,3,10,{7 \over 10},3,4),\\
N &\sim& (1; 1,3, 1,1,-6,8 ), \\
x &\sim &(8;14, 1, 1,-{3 \over 8},0,-{1 \over 2} ), \\
\bar{C} &\sim&( 1;14, 1,\bar{10},{3 \over 10},0,1 ), \\
\bar{S} &\sim& (\bar{36};1, 1, 1,{3 \over 4}
,0,3 ), \\
H &\sim&(1;1, 1, 45,-{3 \over 5},0,0 ).
\end{eqnarray}
and with the superpotential
\begin{equation}
\tilde{W}=Mx\bar{C}\bar{q}+Np\bar{q}+\bar{S}p^2u+\bar{S}x^2+H\bar{C}^2.
\label{superpotential;dual2}
\end{equation}
With the above particle content both gauge factors are asymptotically free.

Next in order to make the theory finite we add an tree-level superpotential
to the original theory,
\be
W= \lambda_1 Q_1 Q_2 Q_3 +
\lambda_2 ( \sum_{i=1}^{10} S \bar{Q}_i \bar{Q}_i). 
\ee
The superpotential $W$ breaks explicitly 
$U(1)_1$ and $U(1)_R$ but leaves $U(1)_2$
unbroken. However a linear combination of $U(1)_1$ and $U(1)_R$ is
preserved. This is an R-symmetry with $Q$, $\bar{Q}$
and $S$ having charges 2/3, 1 and 0. The non-abelian global $SU(3)$
is preserved by the superpotential, the global $SU(10)$ is broken
to its $SO(10)$ subgroup. 
For simplicity we apply the method of \cite{leigh} to show
that this is finite.
Due to the global symmetries
we only have 3 independent gammas, $\gamma_Q$, $ \gamma_{\bar{Q}}$ and
$\gamma_S$. The relevant $\beta$ functions are
\begin{eqnarray}
\beta_{gauge} &\sim& 3 \gamma_q + 10 \gamma_{\bar{Q}} + 5 \gamma_S \\
\beta_{\lambda_1} & \sim& 3 \gamma_q \\
\beta_{\lambda_2} & \sim& \gamma_S + 2 \gamma_{\bar{Q}}
\end{eqnarray}
and hence
\be
\beta_{gauge} \sim \beta_{\lambda_1} + 5 \beta_{\lambda_2}. 
\ee
The conditions of vanishing of all $\beta$ functions
 are hence linearly dependent. Therefore
we obtain a fixed line passing through the origin of coupling space.
Along the fixed line we obtain interacting theories which
are finite in the sense that all $\beta$ functions,
which are the physically relevant quantities,
are zero.

On the dual side $S \bar{Q} \bar{Q}$ becomes the singlet field $H$.
Adding this to the superpotential forces $\bar{C}$ to get a non-zero
vev and breaks the $SO(14)$ to $SO(6)$. On the other hand $Q^3$
is mapped to $\bar{C}^{10} W_{SO(14)}^2$, where $W_{SO(14)}$ denotes
the chiral field strength of the $SO(14)$ gauge group \cite{sakai}.
This term is highly non-renormalizable. Moreover the effect of adding
it to the superpotential is not easy to study. Clearly
the dual theory is far from being finite. It definitely looks much
more complicated than the theory we started with. Nevertheless it
is interesting to see that the dual theory can be obtained.

\subsection{The dual of a realistic $SU(5)$ finite gauge theory}

We have already observed that deconfinement leads to `ugly'
dualities. In addition to the criticism
mentioned above we have seen
that in general some
of the superpotential terms necessary to make the original theory finite are
mapped in the dual theory 
to operators which are quartic or of even higher order leading
to a highly non-renormalizable field theory. Nevertheless it
is interesting to see that using the deconfinement method it is
in principle possible to construct duals even for realistic
models like the one presented in \cite{kmz}.

\subsubsection{The realistic finite unified $SU(5)$ theory}

Let us recall the main features of the
finite unified model based on $SU(5)$. From the classification of
theories with vanishing one-loop
$\beta$ function for the gauge coupling
\cite{HPS}, one can see that
using $SU(5)$ as gauge group there
exist only two candidate models which can
accommodate three fermion
generations. These models contain the chiral supermutiplets
${\bf 5}~,~\overline{\bf 5}~,~{\bf 10}~,
~\overline{\bf 10}~,~{\bf 24}$
with the multiplicities $(6,9,4,1,0)$ and
 $(4,7,3,0,1)$, respectively.
Only the second one contains a ${\bf 24}$-plet which can be used
for spontaneous symmetry breaking (SSB) of $SU(5)$ down
to $SU(3)\times SU(2) \times U(1)$. (For the first model
one has to incorporate another way, such as the Wilson flux
breaking to achieve the desired SSB of $SU(5)$ \cite{kmz}).
Therefore, we would like to concentrate only on the second model.

To simplify the situation, one neglects the intergenerational
mixing among the lepton and quark supermultiplets and consider
the following $SU(5)$ invariant cubic
superpotential for the (second)
model:
\bea
W &=& \sum_{i=1}^{3}\sum_{\alpha=1}^{4}\,[~\frac{1}{2}g_{i\alpha}^{u}
\,{\bf 10}_i
{\bf 10}_i H_{\alpha}+
+g_{i\alpha}^{d}\,{\bf 10}_i \overline{\bf 5}_{i}\,
\overline{H}_{\alpha}~] \nn\\
 & & +\sum_{\alpha=1}^{4}g_{\alpha}^{f}\,H_{\alpha}\,
{\bf 24}\,\overline{H}_{\alpha}+
\frac{g^{\lambda}}{3}\,({\bf 24})^3~,~
\mbox{with}~~g_{i \alpha}^{u,d}=0~\mbox{for}~i\neq \alpha~,
\eea
where the ${\bf 10}_{i}$'s
and $\overline{\bf 5}_{i}$'s are the usual
three generations, and the four
$({\bf 5}+ \overline{\bf 5})$ Higgses are denoted by
 $H_{\alpha}~,~\overline{H}_{\alpha} $.
The superpotential is not the most general one, but
by virtue of the non-renormalization theorem,
this does not contradict the philosophy of
the coupling unification by the reduction
method (a RG invariant fine tuning is a solution
of the reduction equation). In the case at hand,
however, one can
find a discrete symmetry that can be imposed
on the most general cubic superpotential to arrive at the
non-intergenerational mixing \cite{kmz}.  This is given in 
table (\ref{tbl}).

\begin{table*}[t]
\caption{The charges of the $Z_7\times Z_3 \times Z_2$ symmetry}
\label{tbl}
$$
\begin{tabular}{|c|c|c|c|c|c|c|c|c|c|c|c|c|c|c|}
\hline
& ${\bf 10}_1$ & ${\bf 10}_2$ &
${\bf 10}_3$ &
 $\bar {\bf 5}_1$ & $\bar {\bf 5}_2
$ & $\bar {\bf 5}_3$& $H_1$ & $\bar{H}_1$&
  $H_2$ &$\bar{H}_2$& $H_3$ & $\bar{H}_3$ &
  $H_4$ & $H_{24}$ \\ \hline
$Z_7$ &1 & 2  & 4 &  4  &
1  &  2  & 5&-5&  3 &-3& 6&-6 & 0&0\\ \hline
$Z_3$ &1 &2   &0  & 0 &0
&0 &1 &-1& 2 &-2&0& 0 & 0&0\\ \hline
$Z_2$&1&1&1&1&1&1&0&0&0&0&0&0&0&0\\
\hline
\end{tabular}
$$
\end{table*}

Given the superpotential $W$,
 the $\beta$ functions of the model
can be computed and were found to be \cite{georgenew}
\bea
\beta^{(1)}_{g} &=& 0~,\nn\\
\beta^{u(1)}_{i\alpha} &=& \frac{1}{16\pi^2}\,
[\,-\frac{96}{5}\,g^2+
6\,\sum_{\beta=1}^{4}(g_{i\beta}^{u})^{2}+
3\,\sum_{j=1}^{3}(g_{j\alpha}^{u})^{2}
+\frac{24}{5}\,(g^{f}_{\alpha})^{2}\nn\\
&&+4\,\sum_{\beta=1}^{4}(g_{i\beta}^{d})^{2}
\,]\,g_{i\alpha}^{u}~,\nn\\
\beta^{d(1)}_{i\alpha} &=& \frac{1}{16\pi^2}\,
[\,-\frac{84}{5}\,g^2+
3\,\sum_{\beta=1}^{4}(g_{i\beta}^{u})^{2}
+\frac{24}{5}\,(g^{f}_{\alpha})^{2}+
4\,\sum_{j=1}^{3}(g_{j\alpha}^{d})^{2}\nn\\
&&+6\,\sum_{\beta =1}^{4}(g_{i\beta}^{d})^{2}\,]\,g_{i\alpha}^{d}~,\\
\beta^{\lambda(1)} &=& \frac{1}{16\pi^2}\,
[\,-30\,g^2+\frac{63}{5}\,(g^{\lambda})^2+
3\,\,\sum_{\alpha =1}^{4}(g_{ \alpha}^{f})^{2}
\,]\,g^{\lambda}~,\nn\\
\beta^{f(1)}_{\alpha} &=& \frac{1}{16\pi^2}\,
[\,-\frac{98}{5}\,g^2+3\,\sum_{i=1}^{3}(g_{i \alpha}^{u})^{2}
+4\,\sum_{i=1}^{3}(g_{i \alpha}^{d})^{2}
+\frac{48}{5}\,(g^{f}_{\alpha})^{2}\nn\\
&&+\sum_{\beta=1}^{4}(g_{\beta}^{f})^{2}
+\frac{21}{5}\,(g^{\lambda})^{2}
\,]\,g_{\alpha}^{f}~.\nn
\eea
Then regarding the gauge coupling $g$ as the primary
coupling one can solve the reduction equations with
the power series ansatz. It was found that the power series,
\bea
(g_{i i}^{u})^2 &=&\frac{8}{5}g^2+\dots~,
~(g_{i i}^{d})^2 =\frac{6}{5}g^2+\dots~,~
(g^{\lambda})^2=\frac{15}{7}g^2+\dots~,\nn\\
(g^{f}_{4})^2 &=& g^2~,~(g^{f}_{\alpha})^2=0+\dots~~(\alpha=1,2,3)~,
\label{fut-bc}
\eea
exists uniquely,
where $\dots$ indicates higher order terms and
all the other couplings have to vanish.
One can 
verify that the higher order terms can be uniquely
computed.

 Consequently, all the one-loop $\beta$ functions of the theory vanish.
Moreover, all the one-loop anomalous dimensions for the chiral
supermultiplets,
\bea
\gamma^{(1)}_{{\bf 10}i} &=& \frac{1}{16\pi^2}\,
[\,-\frac{36}{5}\,g^2+
3\,\sum_{\beta=1}^{4}(g_{i\beta}^{u})^{2}+
2\,\sum_{\beta=1}^{4}(g_{i\beta}^{d})^{2}
\,]~,\nn\\
\gamma^{(1)}_{\overline{{\bf 5}}i} &=& \frac{1}{16\pi^2}\,
[\,-\frac{24}{5}\,g^2+
4\,\sum_{\beta =1}^{4}(g_{i\beta}^{d})^{2}\,]~,\nn\\
\gamma^{(1)}_{H_{\alpha}} &=& \frac{1}{16\pi^2}\,
[\,-\frac{24}{5}\,g^2+
3\,\,\sum_{i =1}^{3}(g_{i\alpha}^{u})^{2}+
\frac{24}{5}(g_{\alpha}^{f})^2\,]~,\\
\gamma^{(1)}_{\overline{H}_{\alpha}} &=& \frac{1}{16\pi^2}\,
[\,-\frac{24}{5}\,g^2+
4\,\,\sum_{i =1}^{3}(g_{i\alpha}^{d})^{2}+
\frac{24}{5}(g_{\alpha}^{f})^2\,]~,\nn\\
\gamma^{(1)}_{{\bf 24}} &=& \frac{1}{16\pi^2}\,
[\,-10\,g^2+
+\sum_{\alpha=1}^{4}(g_{\alpha}^{f})^{2}
+\frac{21}{5}\,(g^{\lambda})^{2}\,]~,\nn
\eea
also vanish in the reduced system.
As it has already been mentioned in section 3.1, these conditions  are
necessary and sufficient for finiteness to all orders in perturbation
theory.

In most of the previous studies of
the present model \cite{model1,model}, however,
the complete reduction of the Yukawa couplings,
which is necessary for all-order-finiteness,
was ignored.  They have used the freedom
offered by the degeneracy in the one- and two-loop
approximations in order to make
specific ans{\" a}tze that could lead to phenomenologically acceptable
predictions.
In the above model, a diagonal solution for the Yukawa
couplings was found, with each family coupled to a different Higgs.
However, the fact that mass terms
do not influence the RG functions in a certain
class of renormalization schemes was used in order to introduce
appropriate mass terms that permit to perform a rotation in the Higgs
sector such that only one pair of Higgs doublets, coupled to
the third family, remains light and acquires a
non-vanishing vev \cite{model}.
Note that the effective coupling of the Higgs doublets
to the first family after
the rotation is very small avoiding in this way a potential problem
with the proton lifetime \cite{proton}.
Thus, effectively,
we have at low energies the Minimal Supersymmetric Standard Model
(MSSM) with
only one pair of Higgs doublets
satisfying at $M_{\rm GUT}$ the following boundary conditions 
\be
 g_{t}^{2}  = \frac{8}{5} g^2+O(g^4)~,~
 g_{b}^{2}=g_{\tau}^{2}=\frac{6}{5} g^2+O(g^4)~,
\ee
where $g_i$ ($i=t, b, \tau$) are the top, bottom
and tau Yukawa couplings
of the MSSM, and the other Yukawa couplings
should be regarded as free.

The model predicts, among others, the top quark mass $M_t=183 GeV$
subject to corrections of less then 4\%
\cite{percent}.
Another variant of the present model has been suggested in 
\cite{Kobayashi}.
Adding soft
breaking terms (which are supposed not to influence the
$\beta$ functions beyond $M_{\rm GUT}$),
we can obtain supersymmetry breaking.
The conditions on the soft breaking terms to preserve
one-loop finiteness have been given already some time ago
\cite{soft}.
Recently, the same problem
in two-loop orders has been addressed \cite{jj}.
Even more recently a new solution to the two-loop finiteness
conditions was found with very interesting
phenomenological implications \cite{Kobayashi}.
It is an open problem whether there exists a suitable set of conditions
on the soft terms for all-loop finiteness.

\subsubsection{Constructing the dual}

To construct the dual of for the model describe above, we first
consider a theory with the same gauge group and matter content,
but without a tree level superpotential
for the original theory. This model
can be dualized. We will add the superpotential later
in order to make the theory all loop finite and
look for the corresponding effects on the dual side.
Without the superpotential several non-abelian global symmetries
are restored. This simplified model has the following
symmetry group
\be
G=SU(5)_{local} \times  [SU(4)  \times 
SU(7)  \times SU(3) \times U(1)_5 \times U(1)_{10} \times
U(1)_{24}  \times U(1)_R]_{global}
\ee
and  matter content
\begin{eqnarray}
Q &\sim&(5;4,1,1,7,9,10,0),\\
\bar{Q} &\sim&(\bar{5};1,7,1,-4,0,0,1),\\
T &\sim&(10;1,1,3,0,-4,0,1),\\
H &\sim&(24;1,1,1,0,0,-4,\frac{2}{5}).
\end{eqnarray}
Deconfining the adjoint by a $SU(4)$ gauge group we get the following
gauge theory with symmetry group
$$
G_{deconfined}=[SU(5) \times SU(4)]_{local} \times [SU(2)_A \times
SU(2)_B \times SU(2)_C \times SU(4) \times SU(7) \times
$$
\be
\times U(1)_5 \times U(1)_A \times U(1)_B \times U(1)_C \times
 U(1)_R]_{global}
\ee
and matter content
\begin{eqnarray}
Q &\sim&(5;1,1,1,1,4,1,7,3,3,3,0),\\
\bar{Q} &\sim& (\bar{5};1,1,1,1,1,7,-4,0,0,0,1),\\
a&\sim &(5;1,2,1,1,1,1,0,-2,0,0,1/2),\\
\alpha &\sim&(1;1,2,1,1,1,1,0,10,0,0,-1/2)\\
A &\sim&(\bar{5};1,1,1,1,1,1,0,-8,0,0,2),\\
b &\sim&(5;1,1,2,1,1,1,0,0,-2,0,1/2),\\
\beta &\sim&(1;1,1,2,1,1,1,0,0,10,0,-1/2),\\
B &\sim&(\bar{5};1,1,1,1,1,1,0,0,-8,0,2),\\
c &\sim& (5;1,1,1,2,1,1,0,0,0,-2,1/2),\\
\gamma&\sim&(1;1,1,1,2,1,1,0,0,0,10,-1/2),\\
C &\sim&(\bar{5};1,1,1,1,1,1,0,0,0,-8,2),\\
x &\sim&(5;4,1,1,1,1,1,0,0,0,0,1/5),\\
\tilde{x} &\sim&(\bar{5};\bar{4},1,1,1,1,1,0,0,0,0,1/5),\\
p_2 &\sim&(1;1,1,1,1,1,1,0,0,0,0,8/5),\\
p_5 &\sim&(5;1,1,1,1,1,1,0,0,0,0,6/5),\\
\tilde{p_5} &\sim&(\bar{5};1,1,1,1,1,1,0,0,0,0,6/5).\\
\end{eqnarray}
The superpotential in this deconfined theory is
\be
W= a^5 \alpha+ b^5 \beta+ c^5 \gamma +A \alpha a
+B \beta b + C \gamma c+p_2 x \tilde{x}+p_5 x^4  +
\tilde{p_5} \tilde{x}^4  .
\ee
Now we can dualize the $SU(5)$ factor to obtain a dual model. We show
some details of this dual in the appendix. 

Adding the tree-level superpotential required for finiteness both
on the original and the dual side does not pose a problem,
since it is known how the various operators map to each other
under duality. While some of the cubic terms get mapped
to mass terms in the dual theory, many of them
get mapped to baryonic operators, which involve a product of
ten fields. Adding these to the dual superpotential
leaves us with a highly non-renormalizable theory. 

\section{Finite gauge theories and their duals from branes}

Recently it has become clear, that field theory results can
be derived by studying the dynamics of branes in string theory. Here
we will discuss some examples, which have an easy realization
in terms of a brane setup.

One important insight for understanding various duality symmetries
in string theory is that string theory contains a new kind of object, the
Dirichlet (D)-brane \cite{dbrane}, 
which is some kind of topological defect, where open strings can end.
With the help of the D-branes
 many of the connections among dual theories have been clarified.

 More specifically a D-brane is a hypersurface on space time on
which open strings are allowed to end. The open strings are then quantized
in the usual way with the only difference that the end points satisfy
Dirichlet boundary conditions $X^{\mu}(0) = X^{\mu}(\pi) = x^{\mu}$ for the
coordinates normal to the hypersurface, which makes sense for a
hypersurface of any number of dimensions, say p space and one time.
The real novelty of the D-brane appears when several parallel branes are
brought into contact. Then open strings stretching from one D-brane to
another produce new states, that become massless when the D-branes
coincide. The discussion now is exactly as for the type I strings. In fact
the type I open strings can be regarded as associated with 32 D9-branes
filling the space time. The open string fields again become matrices, and
are governed by a dimensionally reduced supersymmetric gauge theory based
on the group U(N). The Lagrangian is
\be
L = 1/g_s TrF^2 + 1/(g_s \alpha^{'2})Tr (DX)^2 + \bar{\psi}\slashD\psi +
1/(g_s\alpha^{'4}) \sum Tr[X^i,X^j]^2.
\ee
Separating the branes in space corresponds to giving  a vev to the matrix
$
X^i$.

 The promotion of space-time coordinates to matrices reminds very strongly
on noncommutative geometry, and the picture has indeed many similarities to
the construction of gauge theories in terms of the noncommutative
geometry of a discrete bundle over space-time
\cite{connes}. Although the
matrix nature of the D-brane coordinates $X^i$ is important in describing
the full dynamics, the moduli space of D-branes in flat space  is
$[X^i,X^j] = 0$. Therefore in low energies, where the moduli space
approximation is a good one, the non-commutativity plays no role.

 During the past year many interesting results, both in field theory and in
string theory, were discovered by studying the worldvolume dynamics of
branes in string theories. A particular construction has been used in
\cite{HW} in order to study $N = 4$ supersymmetric gauge theories in
2 + 1 dimensions. Of particular interest for our purposes is the work in
\cite{EGK}, where the authors following \cite{HW} found a
rather simple description on the way the $N=1$ duality of \cite{seiberg}
arises in 4 dimensions. Later it was realized that many detailed
properties of these field theories can be studied by
lifting the setup to M-theory \cite{wittenM}.
In the following we will discuss some examples of
finite gauge theories which have an easy realization in terms of a brane
setup. However there are two facts that limit the construction of such
models. The first is that a prerequirement in the construction of
finite theories is the existence of a tree order superpotetial. The second
is the fact that all the gauge theories resulting from branes so far are
vector like \footnote{See however \cite{chiral}}.

\subsection{SQCD with one additional superfield in the adjoint}

Let us consider a vector-like gauge theory based on the $SU(N_c)$ gauge
group which contains $N_f$ superfields $Q_i$, $\bar{Q_i}$ transforming as
in SQCD and an additional one $X$ transforming according to the adjoint
representation.

The vanishing of the one-loop $\beta$ function requires that $N_f = 2N_c$.
In order to upgrade the model to an all loop finite one, we have to add
an appropriate superpotential. By adding the superpotential
\be
\label{1adj}
 \lambda Tr X^3 + h \bar{Q_i} X Q_i ,
\ee
the model is derivable from the brane setup of fig. 1.
We consider several branes of IIA string theory in 10 flat dimensions.
In addition to the Dirichlet branes mentiones earlier, there
are also solitonic fivebranes which are called NS fivebranes.
The only property of these NS fivebranes we will need in the
following is that D4 branes can end on them.
In fig. 1 one studies the
worldvolume theory of the $N_c$ D4 branes stretching infinetely
in the 0123 directions and a finite interval between
NS fivebranes in the 6 direction, effectively
compactifying their world volume gauge theory
from 4+1 to 3+1 dimensions. The left NS branes
stretch in the 012345 directions, while the right NS$'$
branes are rotated with respect to this and
stretch in the 012389 directions. The boundary
conditions on the NS and NS$'$ branes project out several
degrees of freedom and break supersymmetry down to $N=1$.
We remain with an $SU(N_c)$ gauge theory with adjoint
fields $X$ and $X'$ coupling via a superpotential
\be
W=\lambda Tr X^{k+1} + \lambda' Tr X'^{k'+1}+h \tilde{Q_i} X Q_i.
\ee
Note that for $k=k'=1$ the superpotential generates masses
for the adjoint fields and the theory reduces to
pure SYM at low energies.
By adding $N_f$ D6 branes stretching in the 0123789 directions
in between the 5 branes we add $N_f$ chiral matter multiplets
in the fundamental and in the anti-fundamental representation.

For $k=2$, $k'=1$ we hence obtain an $SU(N_c)$ gauge
theory with $N_f$ flavours and an additional adjoint field
with the superpotential (\ref{1adj}).
Note that the superpotential (\ref{1adj}) breaks the $SU(N_f)_L \times
SU(N_f)_R$ global symmetry to its diagonal subgroup.

\vspace{1cm}
\fig{Brane configuration corresponding to SQCD with additional adjoint matter
}{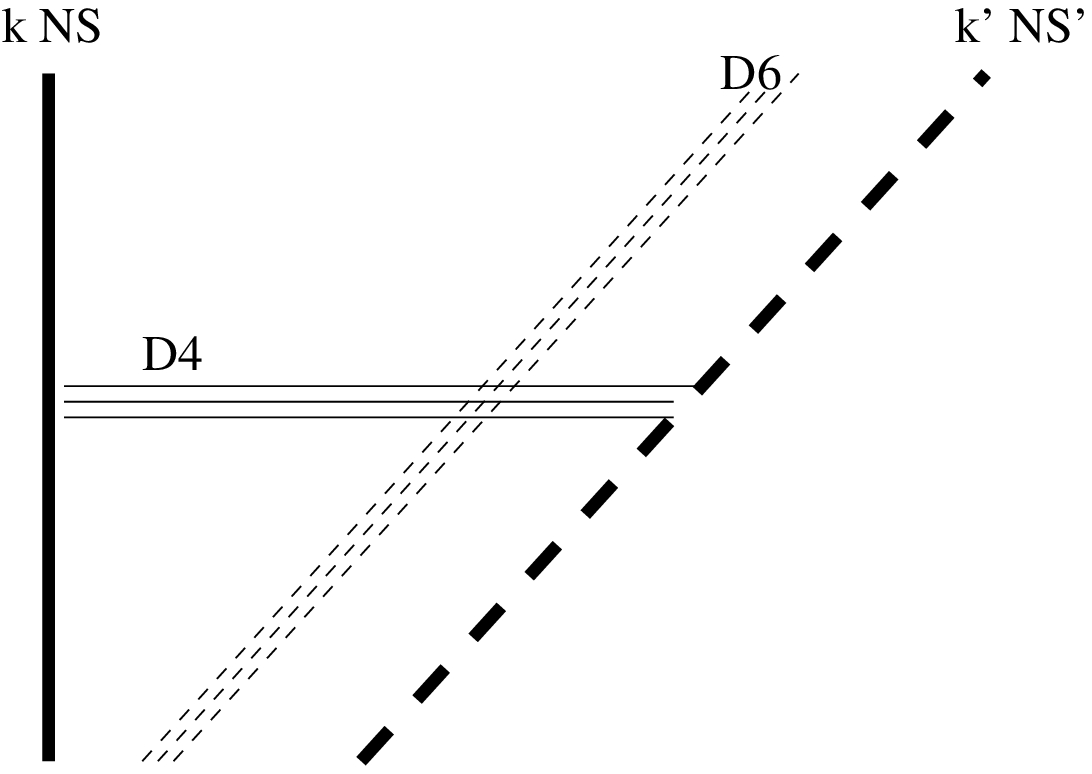}{7truecm}
\vspace{1cm}

 However requiring vanishing of the one-loop anomalous dimensions of the
superfields involved, i.e. $\gamma^{(1)}_Q$, $\gamma^{(1)}_{\bar{Q}}$,
$\gamma^{(1)}_X$, in order to meet the all loop finiteness conditions we
find
\be
 h = g , \; \; \lambda = 0,
\ee
where $g$ is the gauge coupling.

Therefore the requirement of all loop finiteness reduces the model to the
N=2 gauge theory, for which the appropriate dual description is known
\cite{SW}.
Moreover one should note that according to \cite{Kutasov}
the brane configuration with NS and NS$'$ branes only realizes
the situation where the couplings $\lambda$, $\lambda' \rightarrow \infty$.
In order to
adjust the values of the couplings
 one has to rotate the NS$'$ branes in the 5689
plane. To obtain $h=0$ we have to make a $90^o$ rotation and
hence turn the NS$'$ branes into NS branes. This is
known to be the brane setup for the $N=2$ gauge theory \cite{EGK}.

\subsection{SQCD with two additional superfields in the adjoint}

 Our next example is again a vector-like theory based on $SU(N_c)$
containing $N_f$ flavours $Q_i$, $\bar{Q}$ transforming as in SQCD but it
it has in addition two superfields $X$, $Y$ instead of one transforming
according to the adjoint representation.
The vanishing of the one-loop $\beta$ function requires that $N_f = N_c$. Now
in order to make the theory all loop finite we add the following
superpotential
\be
W_e=s_x X^3+s_y Y^3 + \bar{Q} Y Q,
\ee
which again breaks the global flavor symmetry to its diagonal subgroup.

This gauge theory is again derivable from  brane setup
\cite{Kutasov} of fig1, this time with $k=k'=2$.
Moving the NS branes to the right and thereby passing
the D6 and NS$'$ branes one obtains the situation
shwon in fig. 2. Since this move is thought to be irrelevant
for the IR physics, the setup of fig. 2 describes
the dual of the gauge theory we started with.

\vspace{1cm}
\fig{Dual brane configuration. $k'$ NS$'$ fivebranes
are connected by $\tilde{N}_c = k N_f -N_c$ fourbranes to $k$
NS fivebranes on their right, and by k fourbranes to each of $N_f$
D sixbranes on their left}
{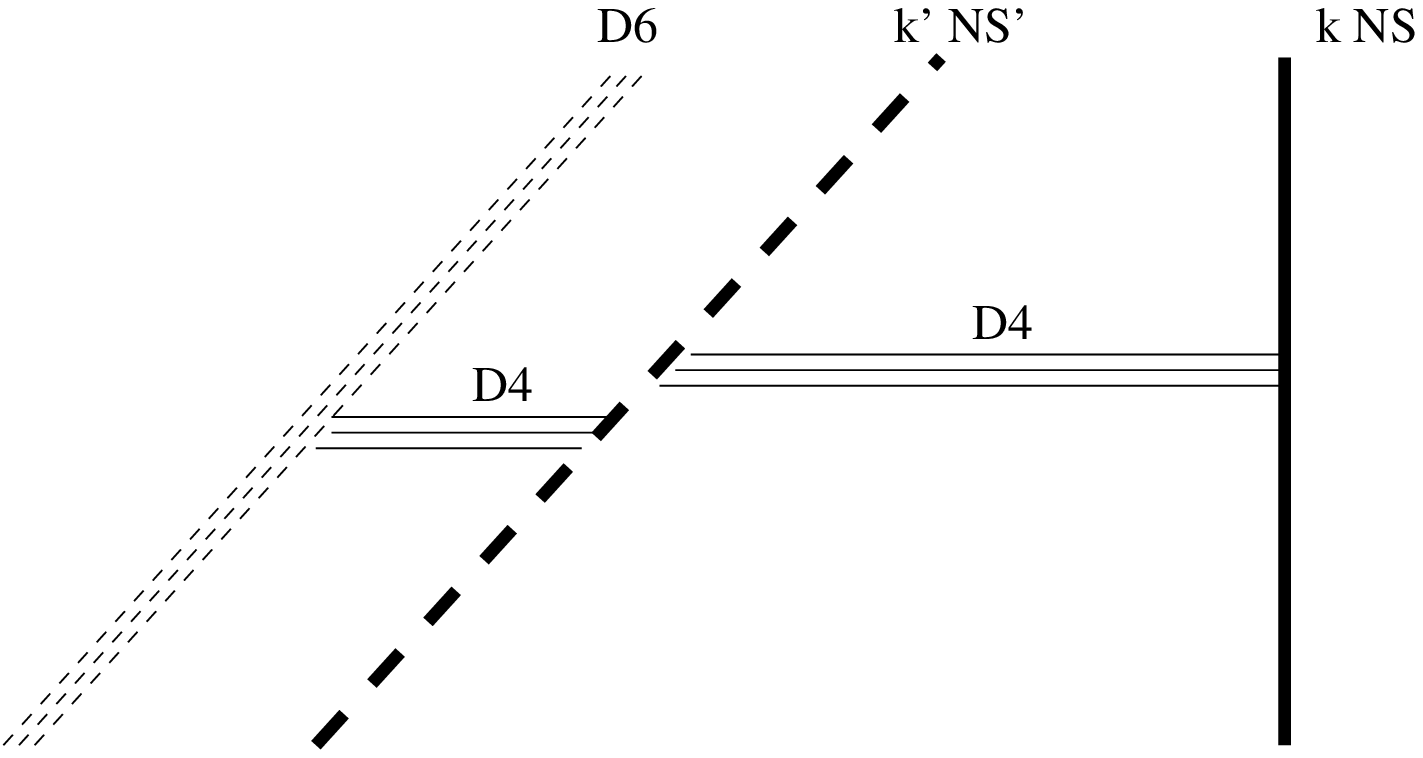}{7truecm}
\vspace{1cm}

The dual is
an $SU(2N_f-N_c)$ gauge theory, again with two adjoints $x$ and $y$ and
$N_f$ fundamentals $q$ and $\bar{q}$. In addition there are gauge singlet
fields $M_1$ and $M_2$ (each one transforming as $adj.+1$ under the
global flavour symmetry). The dual superpotential can be read off from
the branes to be:
\be
\tilde{W}=\tilde{s_x} x^3 +\tilde{s_y} y^3+ \bar{q} y q +M_1 \bar{q} x q+
M_2 \bar{q} q.
\ee
In the potential finite setup with $N_f=N_c$ this is indeed self-dual, but
this time we obtain unwanted quartic operators on the dual side. This
seems rather annoying. Both sides have the right matter content to
be finite, but the 
nonrenormalizable quartic terms in $W$ prevent the dual theory
from being finite.

The solution of this puzzle is deeply connected to the way these dualities
 arise
in the brane picture. As shown by \cite{Kutasov}, the brane only realizes the
configuration where the $s_i\rightarrow \infty$. The appearence
of the quartic terms on the dual side are going to be remnants of this
limit, where new degrees of freedom become massless. This is exactly
what happens in the case of just one adjoint, which we discussed in the
previous section. Here \cite{Kutasov} showed that in addition to
the terms present in the duality found by usual field theory arguments
by \cite{aharony} we get an additional singlet field with a quartic
superpotential coupling in the $s \rightarrow \infty$ limit
seen by the branes.
To achieve a finite theory we have to adjust the $s$. Since they are fixed
to be infinity from the brane point of view, the brane construction will
not help us to understand the dualities of the finite theories. We need
a duality for arbitrary values of the couplings, as is usually obtained
by field theoretic reasoning. Thus we should look for a generalization
of the duality of \cite{aharony} to the 2 adjoint case. Probably
this has to be obtained by the usual `guess and check' procedure. Perhaps
one can deform one of the 2 adjoint dualities
of \cite{Brodie+Strassler} to derive this wanted
duality. This is clearly very important, since this theory seems to
be a very good candidate for a finite, S-dual $N=1$ theory.

\section{Conclusions}

 In the present paper we have been searching for the dual gauge theories
of all-loop finite, $N=1$ supersymmetric gauge theories, following the lines
of \cite{seiberg}. Using established methods for such searching we have been
able to construct the duals of almost all known $N=1$ supersymmetric, chiral
gauge theories (with the exception of $E_6$ models and $SO(10)$
containing antispinors) with vanishing one-loop $\beta$-function. These
theories have first been promoted to all-loop finite ones, by adding
appropriate superpotential and in turn by meeting the requirements of all
loop-finiteness. In addition certain vector-like, all-loop finite, $N=1$
gauge theories and their duals have been discussed, first in the 
standard field theory
framework, and second also using the derivation of
gauge theories from branes. However the brane picture still encounters
several difficulties in the corresponding hunting for finite gauge theories,
as we have seen in the discussion of specific examples.

 From our search, one chiral, $N=1$, all-loop finite gauge theory has been
singled out as a candidate of having S-duality as the $N = 4$ gauge
theories. It is based on the gauge group $SO(10)$ and has matter content
consisting of eight vector and eight spinor superfields. The dual of this
theory is based on the $SU(17) \times Sp(14)$ gauge group. We found
found that there is strong evidence that
after spontaneous symmetry breaking of the first gauge factor to $SU(9)$
the infrared theory is finite.
So, given that both the electric as well as the magnetic theories are finite,
there exist marginal operators, common to both theories, which can be used
to continuously interpolate between the strongly (weakly) coupled
electric theory and the weakly (strongly) coupled magnetic theory, i.e.
there should be $N=1$ S-duality.

 The dual of the realistic finite unified theory \cite{kmz}
based on the $SU(5)$
has been determined and discussed in some detail. However the resulting
dual theory is rather complicated and it does not give, for far, any hint
for a useful use of it. On the other hand, we should note that duals
constructed using  the deconfinement method, as it was the case
for determining the dual of the finite unified $SU(5)$ model, are not
unique. Therefore we cannot exclude the possibility that more interesting
dual theories can be constructed, which also provide in the infrared the
same physics as the original ones.

\vskip1cm
\noindent{\bf Acknowledgements:}

\noindent It is a pleasure to thank J. Kubo for many discussions and for
correspondence concerning the manuscript. We also like to thank 
C. Bachas, S. Ferrara, L. Ibanez and R. Oehme for useful discussions.

\appendix
\section*{Appendix}

Here we present the matter content of the dual
of the realistic $SU(5)$.
For simplicity of notation we 
consider the case without  tree-level
superpotential in the original theory, 
so that we have the full non-abelian global symmetry
group and can organize the fields in a little bit more transparent fashion.

The superpotential are the usual $M q \tilde{q}$ with all the meson
fields appearing with the corresponding quarks.
It is no problem to translate the tree-level superpotential
required for finiteness in this dual language. Some terms
turn out to be mass terms, but many are mapped to baryon
like excitations, which fields to the tenth power and hence
are nonrenormalizable.
The global symmetry structure is broken to the same discrete subgroup
allready present in the original theory. For simplicity
we will only present the matter content of the theory without
superpotential.

Under the symmetry group
\be
\tilde{G}=
[SU(10) \times SU(4) \times SU(2)_A \times SU(2)_B \times SU(2)_C]_{local}
\times [SU(4) \times SU(7)]_{global}\\
\ee
the matter fields transform as
\begin{eqnarray*}
\alpha \sim(1,1,2,1,1;1,1), & \; \;&
\beta \sim(1,1,1,2,1;1,1),\\
\gamma \sim(1,1,1,1,2;1,1), & \; \;&
p_2 \sim(1,1,1,1,1;1,1),\\
q \sim(10,1,1,1,1;\bar{4},1),& \; \;&
\bar{q} \sim(\bar{10},1,1,1,1;1,\bar{7}),\\
a_d \sim(10,1,2,1,1;1,1),& \; \;&
A_d \sim(\bar{10},1,1,1,1;1,1),\\
b_d \sim(10,1,1,2,1;1,1),& \; \;&
B_d \sim(\bar{10},1,1,1,1;1,1),\\
c_d \sim(10,1,1,1,2;1,1),& \; \;&
C_d \sim(\bar{10},1,1,1,1;1,1),\\
x_d \sim(10,\bar{4},1,1,1,1,1),& \; \;&
\tilde{x_d} \sim(\bar{10},4,1,1,1;1,1),\\
p_5^d \sim(10,1,1,1,1;1,1),& \; \;&
\tilde{p_5^d} \sim(\bar{10},1,1,1,1;1,1),\\
M \sim(1,1,1,1,1;4,7);& \; \;&
M_1 \sim(1,1,1,1,1;4,1),\\
M_2 \sim(1,1,1,1,1;4,1),& \; \;&
M_3 \sim(1,1,1,1,1;4,1),\\
M_4 \sim(1,\bar{4},1,1,1;4,1),& \; \;&
M_5 \sim(1,1,1,1,1;4,1),\\
M_6 \sim(1,1,2,1,1;1,7),& \; \;&
M_7 \sim(1,1,2,1,1;1,1),\\
M_8 \sim(1,1,2,1,1;1,1),& \; \;&
M_9 \sim(1,1,2,1,1;1,1),\\
M_{10} \sim(1,\bar{4},2,1,1;1,1)& \; \;&
M_{11}\sim(1,1,2,1,1;1,1)\\
M_{12}\sim(1,1,1,2,1;1,7),& \; \;&
M_{13} \sim(1,1,1,2,1;1,1),\\
M_{14} \sim(1,1,1,2,1;1,1),& \; \;&
M_{15} \sim(1,1,1,2,1;1,1),\\
M_{16} \sim(1,\bar{4},1,2,1;1,1),& \; \;&
M_{17} \sim(1,1,1,2,1;1,1),\\
M_{18} \sim(1,1,1,1,2;1,7),& \; \;&
M_{19} \sim(1,1,1,1,2;1,1),\\
M_{20} \sim(1,1,1,1,2;1,1),& \; \;&
M_{21} \sim(1,1,1,1,2;1,1),\\
M_{22} \sim(1,\bar{4},1,1,2;1,1),& \; \;&
M_{23} \sim(1,1,1,1,2;1,1),\\
M_{24} \sim(1,4,1,1,1;1,7),& \; \;&
M_{25} \sim(1,4,1,1,1;1,1),\\
M_{26} \sim(1,4,1,1,1;1,1),& \; \;&
M_{27} \sim(1,4,1,1,1;1,1),\\
M_{28} \sim(1,1,1,1,1;1,1),& \; \;&
M_{29} \sim(1,4,1,1,1;1,1),\\
M_{30} \sim(1,1,1,1,1;1,7),& \; \;&
M_{31} \sim(1,1,1,1,1;1,1),\\
M_{32} \sim(1,1,1,1,1;1,1),& \; \;&
M_{33} \sim(1,1,1,1,1;1,1),\\
M_{34} \sim(1,\bar{4},1,1,1;1,1),& \; \;&
M_{35} \sim(1,1,1,1,1;1,1),\\
h \sim(1,15,1,1,1;1,1),& \; \;&
\end{eqnarray*}


\begin{thebibliography}{99}

\bibitem{west}
See e.g. D. West,
`Introduction to supersymmetry and supergravity', World Scientific Publishing
Co. (1990), for references and relevant discussion.

\bibitem{kmz} D. Kapetanakis, M. Mondrag{\' o}n and
    G. Zoupanos, {\sl Zeit. f. Phys.} {\bf C60} (1993) 181;
    M. Mondrag{\' o}n and G. Zoupanos, {\sl Nucl. Phys.} {\bf B}
    (Proc. Suppl) {\bf 37C} (1995) 98.

\bi{monol}
C. Montonen and D. Olive, {\it Phys. Lett.} {\bf 72B} (1977).

\bibitem{SW}
N. Seiberg and E. Witten, {\it Nucl. Phys.} {\bf B426}, (1994) 19,
hep-th/9407087 and {\it Nucl. Phys.} {\bf B431} (1994) 484,
 hep-th/9408099.


\bibitem{seiberg}
N. Seiberg, `Electric-Magnetic Duality in Supersymmetric Non-Abelian
Gauge Theories',
{\it Nucl. Phys.} {\bf B435} (1995) 129, hep-th/9411149.


\bibitem{leigh}
R. Leigh and M. Strassler, `Exactly Marginal
Operators and Duality in Four Dimensional N=1
 Supersymmetric
       Gauge Theory', {\it Nucl. Phys.} {\bf B447}
(1995) 95, hep-th/9503121.


\bibitem{stringdual}
A.~Font, L.~Ib\'a\~nez, D.~L\"ust and F.~Quevedo,
  {\it Phys. Lett.} {\bf B 249} (1990) 35;\\
S.--J.~Rey, {\it Phys. Rev.} {\bf D 43} (1991) 526;\\
  A.~Sen, {\it Phys. Lett.} {\bf B 303} (1993) 22, {\bf
B 329} (1994) 217;\\
  J.~Schwarz and A.~Sen,  {\it Nucl. Phys.}
{\bf B 411} (1994) 35, hep-th/9304154;\\
M.J. Duff and R. Khuri, {\it Nucl. Phys.} {\bf B 411} (1994) 473, 
hep-th/9305142;\\
C. M. Hull and P. Townsend, {\it Nucl. Phys. } {\bf B 438} (1995) 109, 
hep-th/9410167;\\
E. Witten, {\it Nucl. Phys.} {\bf B 443} (1995) 85, hep-th/9503124;\\ 
S. Kachru and C. Vafa, Nucl. Phys. {\bf B 450} (1995) 
69, hep-th/9505105;\\
J.H. Schwarz, {\it Phys. Lett.} {\bf B 360} (1995) 13, hep-th/9508143;\\
P. Horava and E. Witten, {\it Nucl. Phys. } {\bf B 460} (1996) 506,
hep-th/9510209.


\bibitem{HW}
A. Hanany and E. Witten, `Type IIB Superstrings,
 BPS Monopoles, And Three-Dimensional Gauge Dynamics'
{\it Nucl.Phys.} {\bf B492} (1997) 152, hep-th/9611230.


  
\bibitem{EGK}
S. Elitzur, A. Giveon, D. Kutasov,
`Branes and N=1 Duality in String Theory',
{\it Phys.Lett.} {\bf B400 } (1997) 269, hep-th/9702014.


\bibitem{Brodie+Strassler}
J. Brodie and M. Strassler, `Patterns of Duality in N=1 SUSY Gauge Theories',
hep-th/9611197.

\bi{TensorKutasov}
D. Kutasov, {\it Phys. Lett.} {\bf B351} (1995) 230, hep-th/9503086;\\
D. Kutasov and A. Schwimmer, {\it Phys. Lett.} {\bf B354}
(1995) 315.

 \bi{so10}
M. Berkooz, P. Cho, P. Kraus, M. Strassler,
`Dual Descriptions of SO(10) SUSY Gauge Theories with Arbitrary Numbers of
       Spinors and Vectors',
hep-th/9705003.  


 \bi{berkooz}
M. Berkooz, `The Dual of Supersymmetric SU(2k) with an Antisymmetric
Tensor and Composite Dualities',
{\it Nucl. Phys.} {\bf B452} (1995) 513, hep-th/9505067.
   


\bi{seirev}
For excellent reviews see e.g. K. Intriligator and N. Seiberg,
hep-th/9509066; \\
M. Peskin, hep-th/9702094.


\bi{ADS}
I. Affleck, M. Dine and N. Seiberg,
{\it Phys. Rev. Lett.} {\bf 51} (1983) 1026;
{\it Nucl. Phys.} {\bf B241} (1984) 493.
\bi{Yank}
T. Taylor, G. Veneziano and S. Yankielowicz,
{\it Nucl. Phys.}, {\bf B218} (1983) 493.

\bi{Se}
N. Seiberg, `Exact Results on the Space of Vacua of Four Dimensional 
SUSY Gauge Theories',
{\it Phys. Rev.} {\bf D49} (1994) 6857, hep-th/9402044.


\bi{hooft}
G. 't Hooft, Proc. Cargese Summer School (Plenum, New York, 1980).



\bi{BaZa} T. Banks and A. Zaks, {\sl Nucl.Phys.} {\bf B149} (1982) 189.


\bi{KuCaSt} J. Kubo, {\sl Phys.Rev.} {\bf D52} (1995) 6475; \\
S.A. Caveny and
P.M. Stevenson, hep-ph/9705319.


\bi{Iwas} Y. Iwasaki, hep-lat/9707019.




\bi{fz-npb87} S. Ferrara and B. Zumino, {\sl Nucl. Phys.} {\bf B87}
(1975) 207.

\bi{ab-theo} S.L. Adler and W.A. Bardeen, {\sl Phys. Rev.} {\bf 182}
(1969) 1517.

\bibitem{pisi} O. Piguet and K. Sibold,  \vspace{-0.2cm}
 Int. J. Mod. Phys. {\bf A1}
               (1986) 913;  Phys. Lett. {\bf B177} (1986) 373.

\bi{sohnius}
see e.g. M. Sohnius, `Supersymmetry and Supergravity 1983',
B. Milewski ed., World Scientific Pub. Co. (1983).


\bibitem{PW}
S. Ferrara and B. Zumino, {\it Nucl. Phys.} {\bf B79} (1974) 413;\\
A.J. Parkes and P.C. West, {\sl Phys. Lett.}
{\bf B138} (1984) 99;
             {\sl Nucl. Phys.} {\bf B256} (1985) 340;\\
             P. West, {\sl Phys. Lett.} {\bf B137} (1984) 371;\\
             D.R.T. Jones and A.J. Parkes, {\sl Phys. Lett.} {\bf B160} (1985)
             267;\\
             D.R.T. Jones and L. Mezinescu, {\sl Phys. Lett.} {\bf B136} (1984)
             242; {\bf B138} (1984) 293;\\
             A.J.~Parkes, {\sl Phys. Lett.} {\bf B156} (1985) 73.
             
             
 \bibitem{nonre} J. Wess and B. Zumino, Phys. Phys. {\bf B49} 52;\\
J. Iliopoulos and B. Zumino, {\sl Nucl. Phys.} {\bf B76} (1974) 310;\\
S. Ferrara, J. Iliopoulos and B. Zumino,
{\sl Nucl. Phys.} {\bf B77} (1974) 413;\\
K. Fujikawa and W. Lang, {\sl Nucl. Phys.} {\bf B88} (1975)
61.            
             

\bibitem{LPS} C. Lucchesi, O. Piguet and K. Sibold,
      Helv. Phys. Acta {\bf 61} (1988) 321.
      
      
      
      
\bi{oehme1}       
R.Oehme, {\sl Phys.Lett.}
{\bf B399} (1997) 67.
      

\bi{Zimmer}
W. Zimmermann, {\sl Com. Math. Phys.}
              {\bf 97} (1985) 211;\\
 R. Oehme and W. Zimmermann {\sl Com. Math. Phys.}
              {\bf 97} (1985) 569; \\
              R. Oehme, K. Sibold and W. Zimmermann,
                {\sl Phys. Lett.} {\bf B147} (1984) 117;
              {\bf B153} (1985) 142;\\
R. Oehme, {\sl Prog. Theor. Phys. Suppl.}
              {\bf 86} (1986) 215 
              
  \bi{pisi-npb196}  O. Piguet and K. Sibold, {\sl Nucl. Phys.} {\bf
  B196} (1982) 428; {\bf B196} (1982) 447.
  
\bi{pisi-book} O. Piguet and K. Sibold, {\sl Renormalized
  Supersymmetry}, Birkh\"auser Boston, 1986.  
  
  
\bibitem{LZ} C. Lucchesi and G. Zoupanos,
{\sl All Order Finiteness in
$N=1$ SYM Theories: Criteria and Applications},
hep-ph/9604216,{\it Fortschr. Phys.} {\bf 45} (1997) 129. 


 \bi{piguet} O. Piguet, {\em Supersymmetry, Ultraviolet Finiteness and
  Grand Unification}, hep-th/9606045.
                      

\bi{ermushev}
          A.V. Ermushev, D.I. Kazakov and O.V. Tarasov,
                 {\sl Nucl. Phys.} {\bf B281} (1987) 72;\\
               D.I. Kazakov, {\sl Mod. Phys. Let.} {\bf A2} (1987) 663;
                 {\sl Phys. Lett.} {\bf B179} (1986) 352;\\
               D.I. Kazakov and I.N. Kondrashuk, {\sl Low-energy
             predictions of Susy GUTs: minimal versus finite model},
             preprint E2-91-393.

\bi{NSVZ}
V. Novikov, M. Shifman, A. Vainshtein and V. Zakharov,
{\it Nucl. Phys.} {\bf B260} (1985) 157;\\
M. Shifman, A. Vainshtein and V. Zakharov, {\it Sov. Phys. Usp.}
{\bf 28} (1985).


\bi{grisaru}
M. Grisaru, W. Siegel and M. Ro\v{c}ec, {\it Nucl. Phys.}
{\bf B159} (1979) 429.

\bi{jackjones}
I. Jack, D. Jones and C. North, {\it Nucl. Phys.} {\bf B473} (1996) 308;
{\it Phys. Lett.} {\bf B386} (1996) 138.




\bibitem{seibergso}
K. Intriligator, N. Seiberg,
`Duality, Monopoles, Dyons, Confinement and Oblique
Confinement in Supersymmetric $SO(N_c)$ Gauge Theories',
{\it Nucl. Phys.} {\bf B444} (1995) 125,
hep-th/9503179.



\bibitem{intrisp}
K. Intriligator, P. Pouliot,
`Exact Superpotentials, Quantum Vacua and Duality in Supersymmetric $SP(N_c)$
       Gauge Theories',
{\it Phys.Lett.}{\bf B353}, (1995) 471, hep-th/9505006.




\bibitem{sakai}
T. Sakai, `Duality in Supersymmetric SU(N) Gauge Theory with 
a Symmetric Tensor', {\it Mod.Phys.Lett.} {\bf A12} (1997) 1025,
hep-th/9701155.


              
 \bibitem{HPS} S. Hamidi, J. Patera and J.H. Schwarz,
              {\sl Phys. Lett.} {\bf B141} (1984) 349;\\
           X.D. Jiang and X.J. Zhou,
        {\sl Phys. Lett.} {\bf B197} (1987) 156; {\bf B216} (1985) 160.
            
              
 \bi{Li}
Ling-Fong Li, {\it Phys. Rev.} {\bf D9} (1974) 1723.             
              
              
       
              
              
 \bi{pouliot}
P. Pouliot, `Chiral Duals of Non-Chiral SUSY Gauge Theories',
{\it Phys. Lett.} {\bf B359} (1995) 108, hep-th/9507018.



\bibitem{acci}
R. Leigh and M. Strassler, `Accidental symmetries and
$N=1$ duality in supersymmetric gauge theory',
{\it Nucl. Phys.} {\bf B496} (1997) 132,
hep-th/9611020;\\
J. Distler and A. Karch, `$N=1$ dualities for exceptional
gauge groups and quantum global symmetries',
to appear in {\it Fortschr. Phys.} {\bf 45} (1997),
hep-th/9611088.





\bi{pouasy}
 P. Pouliot,
`Duality in SUSY $SU(N)$ with an Antisymmetric Tensor',
{ \it Phys. Lett.} {\bf B367} (1996) 151, hep-th/9510148.

 \bi{georgenew}
J. Kubo, M. Mondragon and G. Zoupanos,
{\it Acta Phys. Pol.} {\bf B27} (1997) 3911;
{\it Nucl. Phys. Proc. Suppl.} {\bf 56B} (1997) 281.  


\bibitem{model1} S. Hamidi and J.H.~Schwarz,
              {\sl Phys. Lett.} {\bf B147} (1984) 301;
                D.R.T. Jones and S. Raby,
              {\sl Phys. Lett.} {\bf B143} (1984) 137;
              J.E. Bjorkman,  D.R.T. Jones and S. Raby
              {\sl Nucl. Phys.} {\bf B259} (1985) 503.

         
\bibitem{model}J. Le\'on et al,
              {\sl Phys. Lett.} {\bf B156} (1985) 66.
              
 \bi{proton}
N. Deshpande, Xiao-Gang, He and E. Keith, {\sl Phys. Lett.}
{\bf B332} (1994) 88.             
              
\bi{percent}
J. Kubo, M. Mondragon, M. Olechowski and G. Zoupanos,
{\it Nucl. Phys.} {\bf B479} (1996) 25;\\
K. Yoshioka, `Finite SUSY GUT Revisited', hep-ph/9705449.


\bibitem{Kobayashi}
T. Kobayashi, J. Kubo, M. Mondragon, G. Zoupanos,
`Constraints on Finite Soft Supersymmetry-Breaking Terms',
hep-ph/9707425.

\bi{soft}
D.R.T. Jones, L. Mezincescu and Y.-P. Yao, {\sl Phys. Lett.}
{\bf B148} (1984) 317.

\bibitem{jj}
I. Jack and D.R.T. Jones, {\it Phys.Lett.} {\bf B333} (1994) 372.


\bibitem{dbrane} For reviews see: J. Polchinski, hep-th/9611050;\\ 
C. Bachas, hep-th/9701019;\\
L. Thorlacius, hep-th/9708078.

\bi{connes}
See e.g. A. Connes, `Non commutative geometry', Academic Press 1994;
J. Madore, `An Introduction to Noncommutative Differential Geometry and its
Physical Applications', Cambridge Univ. Press, 1995.


\bi{wittenM}
E. Witten, `Solutions Of Four-Dimensional Field Theories Via M Theory',
hep-th/9703166.

\bibitem{chiral}
 J. Lykken, E. Poppitz, S. Trivedi,
`Chiral Gauge Theories from D-Branes',
hep-th/9708134.

\bibitem{Kutasov}
 S. Elitzur, A. Giveon, D. Kutasov, E. Rabinovici, A. Schwimmer,
 `Brane Dynamics and N=1 Supersymmetric Gauge Theory',
 hep-th/9704104.
 
 
 \bibitem{aharony}
O. Aharony, J. Sonnenschein  and S. Yankielowicz,
 `Flows and Duality Symmetries in N=1 Supersymmetric Gauge Theories',
 {\it Nucl. Phys} {\bf B449} (1995) 509, hep-th/9504113.
 
 

\end{thebibliography}
\end{document}